# Direct observation of ultrafast defect-bound and free exciton dynamics in defect-engineered WS$_2$ monolayers

Tae Gwan Park[1,2], Xufan Li[3], Kyungnam Kang[1], Austin Houston[4], Liam Collins[1], Gerd Duscher[4], David B. Geohegan[4], Christopher M. Rouleau[1], Kai Xiao[1,*], Alexander A. Puretzky[1,*]

[1]Center for Nanophase Materials Sciences, Oak Ridge National Laboratory, Oak Ridge, Tennessee 37831

[2]Center for Integrated Nanotechnologies, Los Alamos National Laboratory, Los Alamos, New Mexico 87544

[3]Honda Research Institute USA Inc., San Jose, California 95134

[4]Department of Materials Science and Engineering, University of Tennessee, Knoxville, Tennessee 37996

*Corresponding author(s) email:  xiaok@ornl.gov and puretzkya@ornl.gov

**ABSTRACT:** Defects in two-dimensional transition metal dichalcogenides (TMDCs) broadly affect their optical and electronic properties. Directly capturing the ultrafast processes of exciton trapping and defect-bound exciton formation is crucial for understanding and advancing defect-mediated optoelectronics and quantum technologies. However, the weak transient optical absorption of defect-bound excitons has limited their experimental observation to date. Here, we report the direct observation of the ultrafast dynamics of defect-bound excitons in monolayer WS$_2$ crystals with a high density of mono-sulfur vacancies (V$_S$) and W-site defect complexes (S$_W$V$_S$) resulting from synthesis by alkali metal halide-assisted chemical vapor deposition. The dynamics of excitons bound to these defects, along with their coherent interactions with free excitons, are elucidated using ultrafast optical spectroscopy. Using above band-edge photoexcitation, we find that both free and defect-bound excitons simultaneously form within 300 fs from hot carrier relaxation. The defect-bound excitons exhibit shorter lifetimes than free excitons, leading to a population difference of the corresponding excitonic states and free exciton trapping within a 1–100 ps window. Band-edge photoexcitation of free and defect-bound exciton states reveals ultrafast interconversion within ~150 fs (comparable to our temporal resolution), indicating possible coherent coupling between these states. We further demonstrate efficient up-conversion of defect-bound excitons to free excitons with photon energies up to ~300 meV below the free exciton resonance. These findings provide insights into the ultrafast dynamics of defect-bound excitons in TMDCs and their coupling with free excitons, which are relevant to defect-engineered optoelectronic, quantum photonic, and valleytronic applications.

Defects in two-dimensional semiconductors, such as transition metal dichalcogenides (TMDCs), significantly influence their optical and electronic properties due to their atomically thin nature and high surface-to-volume ratio.[1-3] Intrinsic point defects and grain boundaries are known to trap charge carriers and excitons, leading to reduced carrier mobility,[4,5] and photoluminescence (PL) quantum yield,[6,7] ultimately limiting the efficiency of pristine TMDCs in device applications.[8] At the same time, defect-bound excitons – *i.e.*, excitons localized at defect sites – exhibit emergent phenomena such as ultralong valley lifetimes,[9] and single-photon emission,[10,11] opening promising pathways for defect-engineering in valleytronics and quantum technology.[2,3,12-14] Consequently, understanding and controlling defect states in TMDCs is of both fundamental and applied interest.

A complete understanding of defect-mediated optoelectronic behavior in TMDCs requires access to the full exciton dynamics, including excitation, relaxation, trapping, transport, and recombination processes. While static optical spectroscopy, such as PL, has provided important insights into defect-bound excitons,[9,15-19] their ultrafast dynamics remain elusive. Ultrafast time-resolved optical spectroscopy, which captures these processes on femtosecond to picosecond timescales, offers a powerful approach to probe the transient behavior of excitons.[13] However, because of the reduced dimensionality and weakened oscillator strength due to the localized nature of defect-bound excitons, these optical transitions are hard to probe by ultrafast optical spectroscopy, which is based on optical absorption and reflection.[15] To overcome this, a high defect density at the intended monolayer crystal location is required. However, defects in TMDCs are usually randomly created during exfoliation or synthesis both in location and number, which make direct observation of defect-bound exciton dynamics challenging. This has led to reliance on indirect interpretations based on changes in free exciton dynamics, which often vary depending on defect type, concentration, or the model assumptions used.

Consequently, a detailed understanding of defect-related ultrafast processes – such as the formation timescales of defect-bound excitons, exciton trapping, and relaxation – remains limited. Previous reports using transient absorption,[20-22] terahertz (THz) spectroscopy,[23-25] and theoretical calculations[26,27] have suggested various defect-related carriers and exciton dynamics over specific timescales, including defect-mediated scattering occurring under 500 fs,[20] fast and slow trapping processes on 0.6–3 ps[20-26] and ~100 ps[21,26] timescales, and recombination over ps–ns timescales.[13] However, these values are often inferred from global carrier and free exciton lifetimes and thus depend heavily on fitting models and assumptions. Recent room temperature observations of unusually strong defect-bound exciton PL emissions within flakes with spatial patterns indicative of high defect densities at specific locations[28] in $WS_2$ monolayers grown using alkali metal-assisted chemical vapor deposition (CVD) methods have motivated us to explore distinct defect-related features in transient absorption spectra, which until now have been very limited[29] and mainly predicted theoretically.[26]

In this work, ultrafast pump-probe microscopy is employed to reveal the ultrafast dynamics of defect-bound excitons that are evident in $WS_2$ monolayers with strong room-temperature photoluminescence resulting from high levels of defects introduced during an alkali metal halide-assisted CVD synthesis. Defects were characterized by combining PL, Raman scattering, Kelvin probe force and electron microscopies. Time-resolved reflection spectroscopy revealed distinct resonances for defect-bound and free excitons, allowing us to track their formation and interconversion dynamics at ultrafast timescales.

By performing the ultrafast optical spectroscopy with a variety of pump wavelengths and a broadband probe, we revealed defect-mediated ultrafast exciton dynamics and estimated their timescales: (1) With above band-edge photoexcitation, we observed an unexpectedly fast defect-bound exciton formation time of ~300 fs, comparable to the free exciton formation time. This suggests that photoexcited hot carriers upon relaxation and cooling form excitons and become bound to the defect states almost simultaneously. Subsequently, a population imbalance emerged due to the difference between free and defect-bound exciton lifetimes, resulting in dynamic exciton trapping within a 1–100 ps time window, consistent with previously reported trapping timescales.[20-26] (2) Band-edge photoexcitation of the free and defect-bound exciton states revealed ultrafast population interconversion (~150 fs, comparable to our temporal resolution) despite their significant energy difference (~80 meV).

Remarkably, free exciton populations were observed even when the pump photon energy was ~300 meV below the free exciton resonance, indicating up-conversion from defect-bound states. This up-conversion efficiency and timescale deviate from existing theoretical models based on phonon-assisted defect scattering.[26] We attribute this fast interconversion to coherent coupling between free and defect-bound excitons, mediated by valley-spin polarizations at the most abundant point defects, such as chalcogenide vacancies.[30-33] These interactions may resemble Dexter-like coupling mechanisms, where spin-conserved excitonic states in opposite valleys interact coherently, akin to the coupling observed between A and B excitons.[34-36] Altogether, our findings provide direct evidence of ultrafast defect-bound exciton formation, interconversion, and coupling with free excitons in $WS_2$ monolayers. These insights reveal new opportunities for tailoring excitonic behavior through defect engineering, with potential applications in optoelectronic, quantum photonic, and valleytronics.

## RESULTS AND DISCUSSION

### Sample synthesis and characterization

Figure 1 presents optical images, photoluminescence (PL) spectra, and PL mapping results for CVD-grown $WS_2$ monolayers on a $Si/SiO_2$ substrate synthesized without (sample #1, Figs. 1a-c) and with the addition of the alkali-metal halide NaBr (sample #2, Figs. 1d-f) in the growth environment. The synthesis methods for these $WS_2$ monolayer crystals are detailed in the Methods section. For sample #1, clean triangular crystals and spatially homogeneous PL emission are observed under 532 nm laser excitation as shown in Figs. 1a and 1b, respectively. The PL spectra measured at different positions from the center to the edge (Fig. 1a, positions 1-3) exhibit a single sharp emission peak centered at 636 nm (1.95 eV), corresponding to the free exciton A emission (X) typically observed in $WS_2$ monolayers[15,17,37] (Fig. 1c). In contrast, sample #2 with a similar triangular shape (Fig. 1d) features a distinct PL mapping with weaker central emission and stronger PL at the edge region of the crystal (Fig. 1e). The PL spectra at different positions (Fig. 1d) show a broad and weak emission at 658 nm (1.88 eV) at the center and a stronger peak at 633 nm (1.96 eV) that dominates towards the edges (Fig. 1f). The higher-energy peak at 1.96 eV can be assigned to free exciton emission while the lower-energy emission at 1.88 eV is attributed to defect-bound excitons (D) in monolayer $WS_2$.

The 80 meV energy difference between the X (1.96 eV) and D (1.88 eV) emission bands exceeds the typical exciton-trion separation (20–30 meV),[18,38,39] ruling out trion contributions as the primary source of the redshifted emission. Instead, the defect-bound exciton emissions is likely due to sulfur vacancies, a common defect in TMDCs.[40,41] Usually, the PL emission of defect-bound excitons in monolayer TMDCs is weak or unobservable at room temperature due to their relatively low binding energy and thermal dissociation.[17-19,42] Strong PL emission of defect-bound excitons has been induced even at room temperature, however, by various techniques such as enhancing defect localization through strain or dielectric engineering, or increasing defect density, or using localized optical excitation by near-field resonance excitation.[16,28,43] Previous reports indicate an energy difference of ~100 meV between free and defect-bound excitons in $WS_2$ at room temperature,[16,43] closely matching our observation.

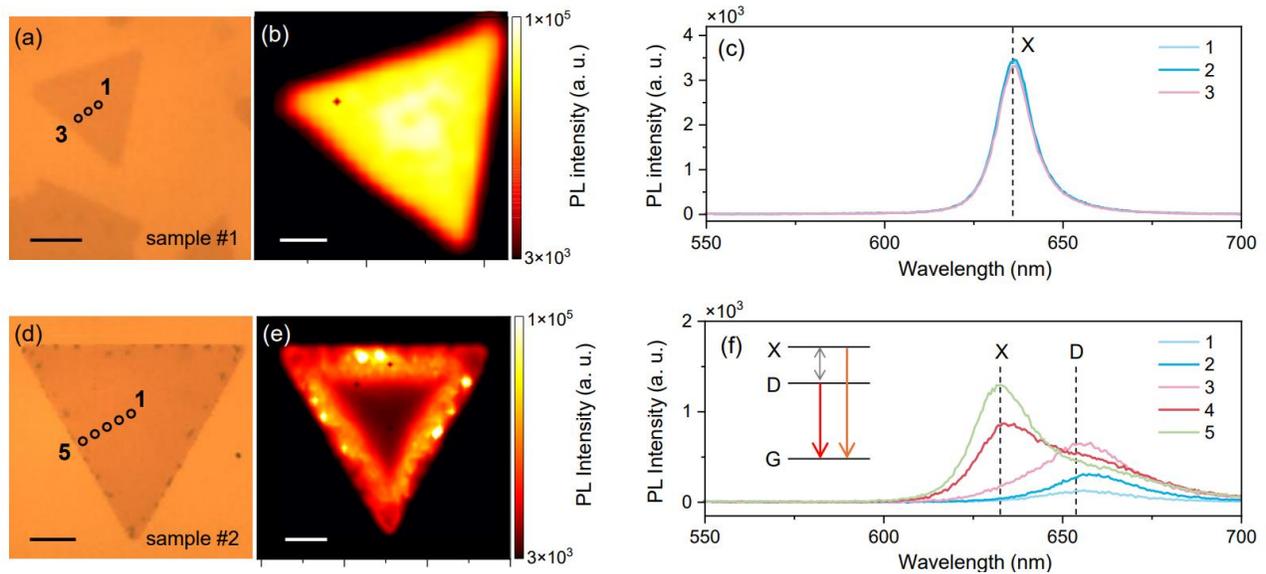

**Figure 1. Defect-bound excitons in alkali-assisted CVD grown $WS_2$ monolayer**. (a) An optical image of CVD-grown $WS_2$ monolayer synthesized without NaBr (sample #1). (b) PL map of sample #1 composed by integrating the PL spectra from 620 nm to 650 nm at each 0.5 μm step. (c) PL spectra acquired at positions 1 to 3 as marked by circles in (a). The vertical dashed line indicates PL center wavelength of the free exciton (X). (d) Optical image of NaBr assisted CVD-grown $WS_2$ monolayer (sample #2). (e) PL map obtained by integrating the spectra at each point from 600 nm to 700 nm (0.5 μm step size). (f) PL spectra measured at the positions 1 to 5 as indicated by circles in (d). The vertical dashed lines indicate the free exciton (X) and defect-bound exciton (D) emission center wavelengths. 532 nm excitation was used for the PL in both cases.

To further examine the defect-bound exciton and its spatial distribution in sample #2, we compared the PL map with the Raman map, which provides rich information about doping, strain, and defects. The PL intensity ratio of the defect-bound excitons to the free excitons (D/X) reveals distinct regions: the center dominated by D emission and the exterior dominated by X emission (Fig. 2a). The edges also exhibit strong D emission, possibly due to edge passivation from residual NaBr clusters.[44]

Figure 2b shows Raman spectra from the center and exterior regions marked in Fig. 2a. The characteristic Raman peaks of $WS_2$[37] – i.e., the LA (173 cm$^{-1}$), 2LA and $E'_2$ (~350 cm$^{-1}$), and $A'_1$ (415 cm$^{-1}$) modes – are identified. The $A'_1$ mode is highly sensitive to electron doping while the $E'_2$ mode is more strain sensitive.[45-47] The $A'_1$ mode peak frequency map (Fig. 2c) shows minimal frequency variation, indicating uniform electron doping across the flake. However, the $A'_1$ intensity map (Fig. 2d) displays a faint boundary similar to the PL map.

Among the primary Raman peaks, the LA phonon intensity map (Fig. 2e) correlates best with the PL emission maps. The intensity ratio of LA to $A'_1$ (LA/$A'_1$) is related to disorder and is proportional to the defect density.[48] The LA/$A'_1$ intensity ratio most closely correlates with the PL map as shown in Fig. 2f. The ratio is ~2.0 at the center and ~1.2 near the exterior, indicating about ~70% higher defect density in the center. A statistical analysis (Supplementary

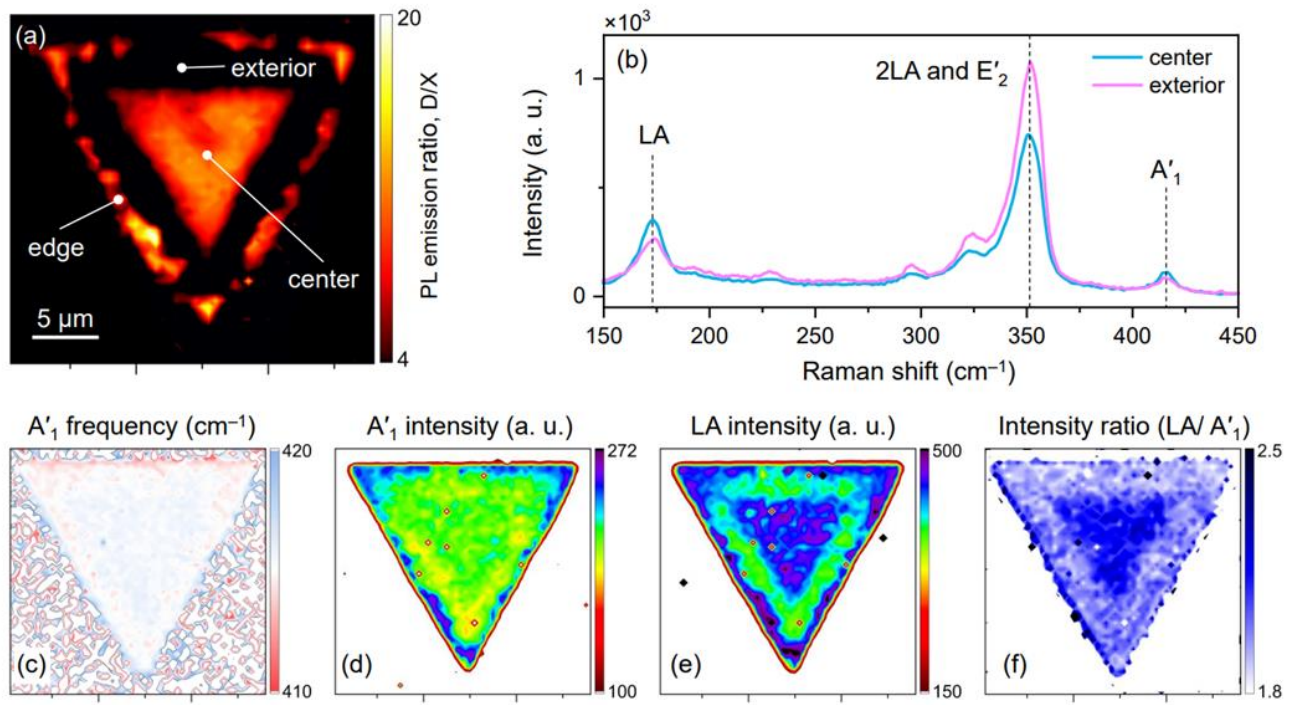

**Figure 2. Correlation of defect-bound exciton emission and Raman characteristics.** (a) D to X ratio PL map of $WS_2$ monolayer synthesized with NaBr (sample #2). The PL is integrated from 640 nm to 675 nm for the D and from 620 nm to 635 nm for the X band (see Fig. 1f). (b) Raman spectra measured at two distinct positions – center and exterior of the monolayer as indicated in (a). (c-f) Raman maps of $A'_1$ phonon frequency (c) and intensity (d), LA phonon band intensity (e), and ratio of LA to $A'_1$ band intensities (f).

Information Fig. S1) confirms that the LA/A′$_1$ intensity ratio is 2.7 times lower in the WS$_2$ monolayer synthesized without NaBr (sample #1) compared to that in the center of sample #2, indicating significantly lower defect density in sample #1 compared to sample #2.

A Kelvin probe force microscopy (KPFM) image also shows a pattern consistent with that observed in PL and Raman mapping, displaying about 0.2 eV lower work function at the center of sample #2 (Fig. S2). The observed lower work function at high defect density position could be related to defect states in the band gap, resulting in the shift of Fermi level toward the conduction band.[49]

To investigate the defect density and spatial distribution within individual WS$_2$ monolayer crystals, we performed atomic resolution scanning transmission electron microscope high angle annular dark field (STEM-HAADF) imaging. WS$_2$ triangle monolayer flakes were transferred onto a QUANTIFOIL gold grid for characterization (Fig. S3). STEM images acquired from both the center and edge regions of the flake revealed two dominant defect types: mono-sulfur vacancies (V$_S$) and W-site defect complexes (S$_W$V$_S$) as shown in Fig. 3a. The S$_W$V$_S$ defect consists of a sulfur atom substituting for a neighboring tungsten atom, accompanied by a sulfur vacancy on a nearby sulfur site. These defect structures were confirmed by comparing experimental ADF images with simulated STEM images (Figs. 3b-d). Averaged cropped images of V$_S$ (Fig. 3e) and S$_W$V$_S$ (Fig. 3f) show distinct lattice distortions. Line profiles (Figs. 3g-h) show that while V$_S$ does not significantly alter the surrounding lattices, S$_W$V$_S$ defects cause a noticeable positive displacement at W – S$_2$ and S$_W$ – V$_S$ sites.

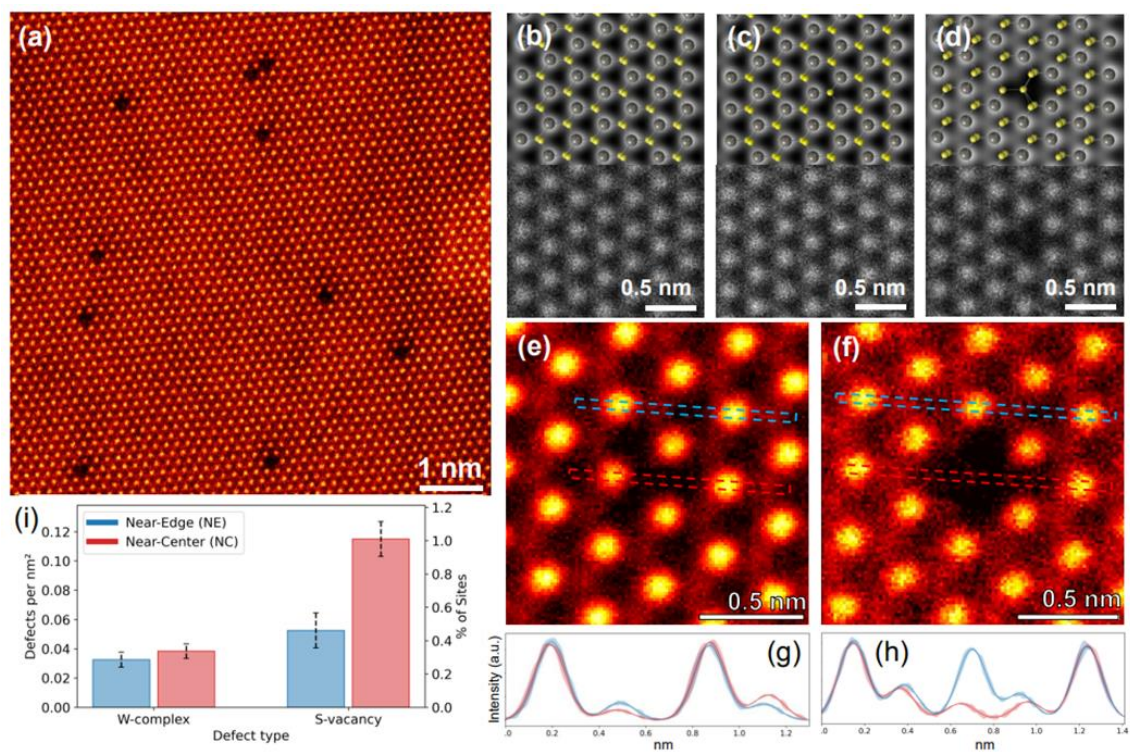

**Figure 3. Identification of defects in monolayer WS$_2$.** (a) Representative high magnification HAADF image from the center region of NaBr-grown WS$_2$ monolayer. Mono-sulfur vacancies (V$_S$) and one antisite defect plus one neighboring sulfur vacancy complex (S$_W$V$_S$) were marked by blue circles and yellow squares, respectively. (b-d) Simulated HAADF-STEM images of the predominant pristine and defective WS$_2$ atomic configurations at high electron dose (top) and low electron dose (bottom). (b) Pristine WS$_2$. (c) WS$_2$ with a single S-vacancy. (d) WS$_2$ with a S$_W$V$_S$-complex. (e) Average of 20 experimental HAADF-STEM images of a S-vacancy found in (a). (f) Average of 10 experimental HAADF-STEM images of a S$_W$V$_S$-complex found in (a). (g) Line profiles acquired along the dashed lines in (e) with standard deviation shaded. (h) Line profiles acquired along the dashed line in (f) with standard deviation shaded. (i) Calculated defect density of V$_S$ and S$_W$V$_S$ by atom-counting of over 10000 atoms near the edge and the center of the WS$_2$ flake (exact regions are shown in the Fig. S3). The error bar means standard deviation (SD) of defect density.

Comparing defect distributions, we found that the center region of the flake exhibits nearly double the $V_S$ density relative to the edge while the density of $S_WV_S$ defects remains similar across both regions (Fig. 3i). This uniform distribution of $S_WV_S$ defects across the flakes allows us to rule out their contribution to the defect-bound exciton emission observed predominantly in the center region. Instead, $V_S$ defects should be the primary contributors to the defect-bound exciton emission. Notably, the $V_S$ density (0.115 nm$^{-2}$) in sample #2 is significantly higher than that of sample 1 (Fig. S4), and falls within the reported range (0.1-10 nm$^{-2}$)[43] necessary for observing defect-bound excitons at room temperature.

By combining PL, Raman, KPFM, and atomic-resolution STEM analysis, we conclude that sulfur vacancies are highly concentrated at the center of the flake, correlating directly with defect-bound exciton emission. Understanding the spatial distribution of defects during monolayer crystal growth remains an open question. However, this work leverages defective WS$_2$ crystals to investigate ultrafast defect-bound exciton dynamics and their coupling with free excitons.

**Ultrafast pump-probe spectroscopy**

Using these unique defective WS$_2$ monolayer crystals, and an optical microscope to provide spatially-resolved (~5 μm resolution) investigation of different regions of individual crystals, we performed ultrafast pump-probe spectroscopy under various photoexcitation conditions to investigate defect-mediated quasi-particle relaxation processes. These conditions include: (1) above band-edge excitation above bandgap ($E_{pump} > E_{gap}$) with a 400 nm pump wavelength, (2) X band-edge excitation to the free exciton transition ($E_{pump} \sim E_X$), and (3) D band-edge excitation to defect-bound exciton transition ($E_{pump} \sim E_D$). We measured the transient reflectivity spectra induced by a pump laser using a white-light continuum as a probe: $\Delta R/R_0 = (R_{pump} - R_0)/R_0$, where $R_{pump}$ and $R_0$ are the reflectivity with and without the pump, respectively.

Figure 4 presents the transient reflectivity results for two different monolayer WS$_2$ crystals: one grown without NaBr (sample #1) and one with a NaBr precursor (sample #2), both at a pump fluence of 66 μJ/cm$^2$. In sample #1 (Fig. 4a), a strong positive peak is observed at the free exciton wavelength around 630 nm, matching with the exciton peak in the linear reflectance and absorption spectrum observed around 630 nm (Fig. S5), meaning the positive $\Delta R/R_0$ band is corresponding to the photobleaching of the X resonance, thereby enabling its use to explore its ultrafast dynamics. Additionally, the $\Delta R/R_0$ spectrum captures the B exciton around 500 nm.

As shown in Fig. 4b, the ultrafast population and decay dynamics of both X and D resonances in sample #2 are clearly captured. Prominent positive peaks are observed in the transient spectrum around 610 nm and 660 nm, due to the absorption bleaching of the X and D bands, which corresponds to the bands around 610 nm and 660 nm in the linear reflectance and absorption spectra (Fig. S5). We also performed direct optical absorption measurements using a sample synthesized on a transparent quartz substrate (Fig. S6), which showed a sharp X peak and a broad, less intense D band at wavelengths close to the X and D bands observed in WS$_2$ on a SiO$_2$/Si substrate (Fig. S5). Notably, the maximum $\Delta R/R_0$ at X in sample #2 is only about 0.3% compared to 3% for that of sample #1 under the same excitation conditions. This 10-fold difference in signal ratio is likely not due to the difference in absorption at the 400 nm pump wavelength, as a previous study has shown no significant variance in absorption coefficient at this wavelength.[50] This suggests that photocarriers with high excess energy from the above band-edge excitation contribute less to the resonance change of the X state in the presence of defects, as is described in detail below.

As the time delay increases, a slight blueshift of approximately 4 nm in the $\Delta R/R_0$ peak for the X band is observed as shown in Figs. 4a and 4b, and more clearly in Figs. S7a and S7b. The generated hot carriers in WS$_2$ by above-bandgap excitation result in a complex $\Delta R/R_0$ spectrum with peak wavelength shifts, broadening, and bleaching of the X resonances. This behavior is attributed to plasma-induced bandgap renormalization and screening of the exciton binding energy,[51-54] along with state-filling[55] and biexciton formation.[56] Despite these complex resonance changes, our primary focus is on population and decay dynamics of free and defect-bound excitons. Therefore, we used the positive ΔR/R$_0$ peaks at the X and D bands as markers for relevant ultrafast dynamics in this study. In Fig. 4c, the positive ΔR/R$_0$ peak values at the X and D bands are plotted at their peak wavelengths for each time delay, with time-dependent center wavelengths shown in Fig. S7c. No notable shifts were observed in the D band (Fig. S7c).

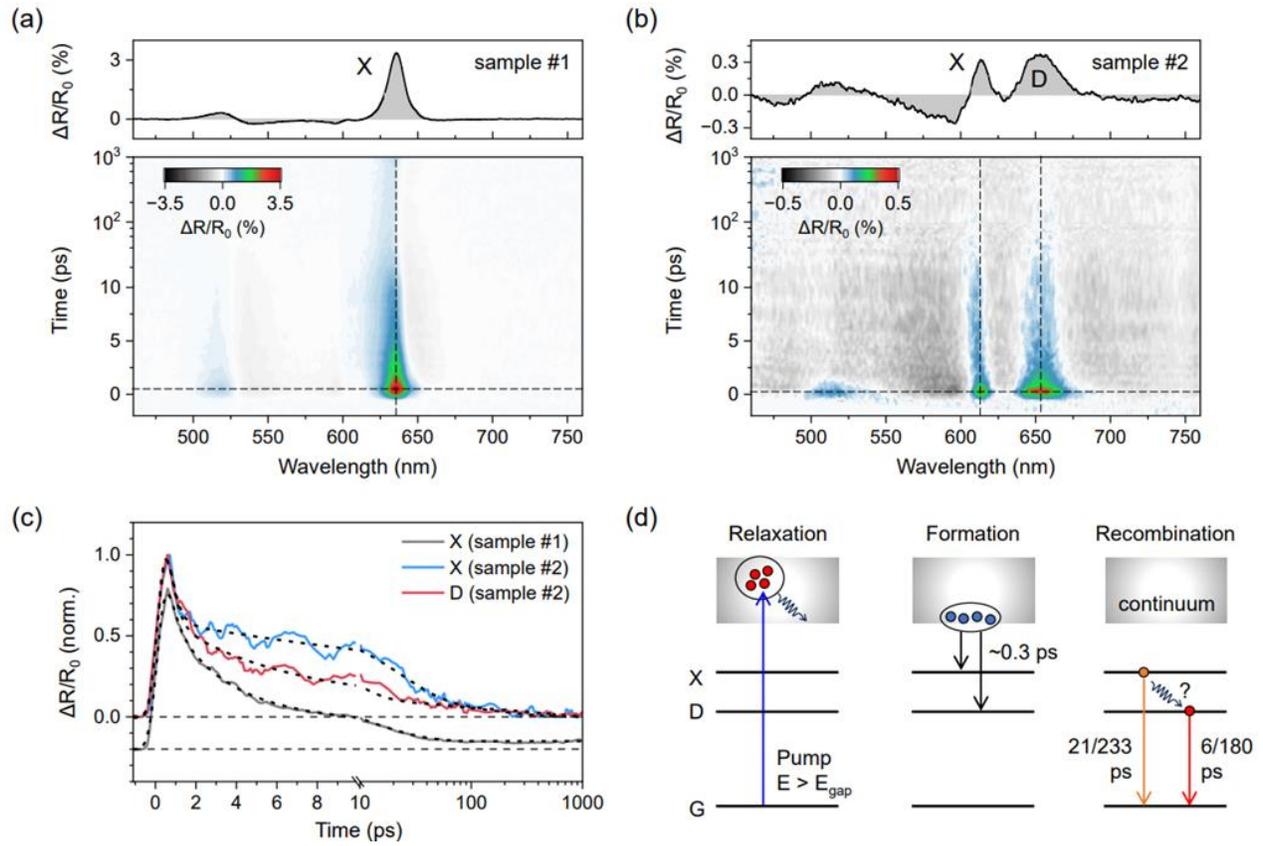

**Figure 4. Transient optical response of WS$_2$ monolayers under above band-edge photoexcitation.** (a,b) Transient reflectivity maps of WS$_2$, as a function of delay and probe wavelength in sample #1(a) and sample #2 (b) under photoexcitation ($\lambda_{ext}$ = 400 nm, pump fluence of 36.4 μJ/cm$^2$). Top panels show extracted ΔR/R spectra at 150 fs for (a) and (b), respectively. (c) Temporal evolution of the excitonic resonances of free excitons (X) (samples #1 and #2) and defect-bound exciton (D) (sample #2). The traces are normalized to their maximum amplitude. A constant vertical offset (−0.2) is applied to the X (sample #1) trace for clarity. The baselines are indicated by the horizontal dashed black lines. (d) Schematic of defect-mediated ultrafast processes showing carrier relaxation, exciton formation, and exciton recombination under above band-edge photoexcitation.

The ultrafast dynamics of the X and D were characterized using a rise time ($\tau_{rise}$) and three exponential decay times ($\tau_1, \tau_2, \tau_3$). Here, the rise time, $\tau_{rise}$, was simply estimated from the time it took for the ΔR/R$_0$ signal to go from 10% to 90% at the X and D transition energies. For above band-edge excitation, free electron–hole pairs are created with high kinetic energies far above the exciton energy. Carrier thermalization and subsequent energy relaxation drive carriers from high to low energy states, increasing the differential reflection signal.[57] These processes contribute to the sub-ps rise kinetics of the transient signal[21,58] while the following fast initial decay ($\tau_1$) corresponds to exciton formation from the cooled carriers.[57,58] Accordingly, $\tau_{rise}$ corresponds to the thermalization and energy relaxation process of the hot carrier.[21,59] We observed a similar $\tau_{rise}$ of approximately 300 fs in the X band in both samples #1 and #2, as well as for the D band in sample #2, consistent with previous reports.[21,59] This suggests that the hot carrier energy relaxation process is mainly driven by carrier-carrier and carrier-phonon scattering. For sample #1, $\tau_1$ was measured to be 0.47 ps, which is in good agreement with previous observations[57,58]. Smaller $\tau_1$ values were observed for both X and D bands in sample #2, measured at 0.3 ps, respectively. The presence of defects in the lattice likely weakens the phonon-assisted scattering process, resulting in a mixture of excitons and unbound electron–hole pairs persisting for longer periods[58] and photoexcited electron-hole pairs becoming strongly localized in trap states before reaching excitonic states. This reduction in X population may lead to the faster $\tau_1$, similar to previous observations in defective WSe$_2$ monolayers.[60] Notably, the observed defect-bound exciton formation time ($\tau_1$ for D) in sample #2 was also ~ 0.3 ps, which is faster than previously reported photocarrier trapping times to range from 1–100 ps[13,21,61-63]. As shown in Fig. 4d, this suggests that photoexcited carriers generated by above-bandgap excitation populate both the X and D states simultaneously after energy relaxation.

The subsequent decay time, $\tau_2$, was measured to be 18 ps for the X band in sample #1. The slow component $\tau_3$ for X kinetics in sample #1 could not be defined by fitting due to the unrelaxed signal persisting beyond 100 ps but is expected to exceed the ns timescale. In sample #2, $\tau_2$ and $\tau_3$ were 21 and 233 ps for the X band, and 6 and 180 ps for the D band, respectively. The larger $\tau_2$ for X (21 ps) in sample #2 compared to that of sample #1 (18 ps) could be attributed to the defect-affected nonradiative recombination process.[13,29] For $\tau_3$, the decay process for the X band in sample #2 (233 ps) is much faster compared to that of sample #1 (> ns), likely due to defect-assisted exciton recombination[13] and slow carrier capture.[21,26] The $\tau_2$ and $\tau_3$ values in the D dynamics (sample #2) reflect fast and slow recombination, driven by the efficient recombination of electron-hole pairs without the momentum mismatch restriction typically present in defect states. The significantly smaller $\tau_2$ and $\tau_3$ for the D band compared to the X in sample #2 indicate that the faster decay of D population may promote additional exciton capture, which can continue to influence the X dynamics on longer timescales (1–100 ps). This aligns well with previously observed timescales of defect-mediated exciton dynamics in the 1 to 100 ps range.[20-26] The observed population and decay kinetics of the X and D states, which yield spectrally distinguishable signals, demonstrate that photocarriers generated via above band-edge excitation rapidly form defect-bound excitons on a similar timescale (~0.3 ps) compared to that of free exciton formation. However, the similar dynamics of both (X and D in sample #2) populations complicate the precise observation of state-to-state coupling and transfer dynamics between these states.

To further investigate exciton transfer from X to D states, we performed pump-wavelength tuning to band-edge excitation conditions for the X band as shown in Fig. 5a. Figure 5b shows results of transient reflectivity measurements with X band-edge photoexcitation at 595 nm, slightly shorter than the X resonance (612 nm) to avoid direct excitation of the D band due to the relatively large bandwidth of the pump ($\Delta\lambda$ ~20 nm) as indicated by the pump scattering in $\Delta R/R_0$ spectrum. The spectrum of the pump laser is shown in Fig. S8. Upon X band-edge excitation, two strong peaks at 613 nm and 648 nm corresponding to the X and D bands were observed. These wavelengths were similar to those observed with 400 nm photoexcitation. However, unlike the results observed with 400 nm photoexcitation, no significant peak shifts of the X and D resonances in the $\Delta R/R_0$ spectrum were detected within the 4 ps temporal window shown in Fig. S9, suggesting that the signal is dominated by state filling.[55] Figures 5c and 5d show the temporal evolution of the X and D bands at their respective peak wavelengths.

The rise time of the X resonance was measured to be ~120 fs, which is shorter than the ~300 fs rise time observed with 400 nm excitation. The absence of a peak shift in the excitonic resonance and the shorter rise time further support that the signal is primarily due to state filling rather than band structure renormalization caused by hot carrier relaxation during 400 nm excitation discussed above. When the pump photon energy is close to the band gap, direct exciton formation occurs in ~10 fs timescale,[55] consistent with geminate exciton formation.[64] The observed 120 fs rise time of X population by X band-edge excitation is therefore limited by the instrumental response and is consistent with the cross-correlation function obtained from cross-phase modulation signals (Fig. S10). Interestingly, a similarly rapid rise time of 142 fs was observed for the D resonance, comparable to the temporal resolution. This rapid signal is attributed to D formation via carrier capture at defect states from exciton states.

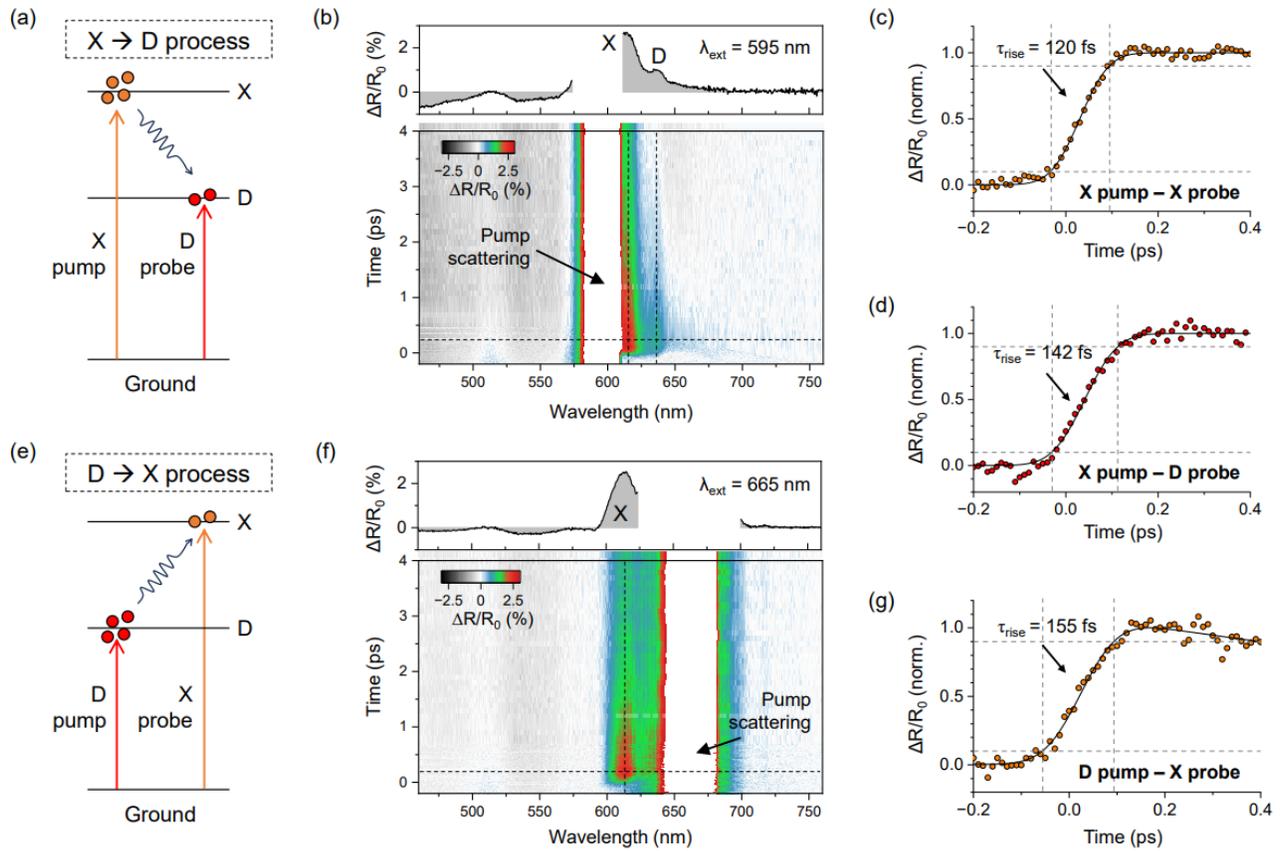

**Figure 5. Transient optical response of WS$_2$ monolayers under band-edge excitation.** (a) Schematic of pump at the X band and probe at the D band. (b) Transient reflectivity maps as a function of delay between pump and probe and probe wavelength following X band-edge excitation ($\lambda_{ext}$ = 595 nm, pump fluence of 61.1 μJ/cm$^2$). Top panels in (b) show extracted ΔR/R$_0$ spectra at 150 fs delay as indicated by the horizontal dashed line. (c, d) Temporal evolution of excitonic population under band-edge excitation: X pump and X probe (c), X pump and D probe (d). (e) Schematic of pump at the D and probe at the X bands. (f) Transient reflectivity maps under $\lambda_{ext}$ = 665 nm pump with pump fluence of 54.7 μJ/cm$^2$. Top panels in (f) shows extracted ΔR/R$_0$ spectra at 150 fs delay as indicated by the horizontal dashed line. (g) Temporal evolution of the X-band population following D band-edge excitation. The vertical lines in (c,d and g) mark the 10 to 90% risetime.

The D formation time is extremely short (142 fs), for example, and it is significantly shorter than the ~1 ps defect-bound exciton formation time observed in the mid-gap defects of MoSe$_2$ using near-infrared pump-probe spectroscopy.[22] However, in MoSe$_2$ the corresponding defect-bound state is located 300 meV below the free exciton state[22] whereas in our WS$_2$ case the energy difference between the D and X states is approximately 80 meV, indicating that the defect-bound state is shallower. This shallower defect state results in a shorter carrier capture time, which is consistent with recent measurement of ultrafast defect capturing time of ~70-80 fs in defective MoS$_2$ monolayer with ~60 meV energy difference between X and D states.[29] Ultrafast exciton trapping has been linked to strong coherent phonon generation at shallow trap states induced by localized defects[65] and our experimental results support this phenomenon. This suggests that a Stokes scattering process could occur between the X and D states, accompanied by strong phonon emission to conserve energy and momentum.

When localized defect potentials are present, momentum is not a good quantum number. Therefore, ultrafast conversion can be expected as we observed. An exciton-to-trion conversion process, which must satisfy energy-momentum conservation conditions, can serve as a good example. In this case, an ~2 ps timescale has been reported for neutral (free exciton) and charged exciton (trion) phonon-mediated incoherent conversion in MoSe$_2$ monolayers.[66] Despite the larger energy difference between the X and D states in our case (~80 meV) compared to the exciton and trion in MoSe$_2$ monolayers (~30 meV), the much faster X → D conversion that we observed may be attributed to a variety of factors such as localization of excitons by defect potentials, relaxing momentum conservation constraints, and enhancing the efficiency of the conversion process.

Given the efficient coupling between X and D states, we anticipated the possibility of up-conversion from the defect-bound state (D) to the free exciton state (X) as shown in Fig. 5e. To explore this possibility, we measured up-conversion PL spectra in the range of photon energies above the excitation photon energy using tunable continuous-wave laser excitation of near D band (648 nm, see Methods) and clearly observed up-converted PL spectra of the X as shown in Fig. S11. The up-converted PL spectrum shows peak at 613 nm corresponding to the X resonance, which is ~ 109 meV higher in photon energy compared to the excitation photon energy.

To further understand the up-conversion dynamics, we performed transient reflectivity measurements with D band-edge excitation as shown in Fig. 5f. The pump excitation had a center wavelength of 665 nm, larger than the D resonance to avoid direct excitation of the X band due to the relatively large bandwidth of the pump laser (Fig. S8). Similar to the up-conversion PL measurements, upon excitation of the D band, the population of the X band was observed at a peak position of 612 nm as shown in Fig. 5f. The X band population feature could be originated from transient bandgap renormalization, caused by photoexcited carriers.[67] However, because D band-edge excitation lies below the bandgap and cannot generate free carriers, photocarrier-induced BGR is unlikely. The symmetric ΔR/R$_0$ spectrum around the X band without a peak shift (Fig. S12) indicates that the X-band bleaching under D band-edge excitation is dominated by state-filling[55] rather than transient bandgap renormalization.[67]

Remarkably, the rise time of the excited X population was measured to be 155 fs, which is comparable to our temporal resolution. It is important to note that the X state is ~80 meV higher in energy than the D state, while the average thermal energy at room temperature is only $k_B T$ ~ 26 meV (where $k_B$ is the Boltzmann's constant and T is room temperature). Consequently, the observed D → X up-conversion is energetically unfavorable from a thermal activation perspective. Therefore, the up-conversion could be attributed to: (1) two-photon absorption (TPA), which leads to direct excitation to a higher energy state,[68] (2) phonon-mediated up-conversion,[15,66,69] or (3) coherent coupling between the X and D states.[34,66,70,71]

In the case of TPA where $2E_{pump} > E_{gap}$, a slower rise time would be expected, similar to the results seen with the 400 nm pump (Fig. 4c). We also measured the dependence of the transient reflection on the pump fluence ($F_{pump}$) to verify the possibility of TPA as the mechanism behind the D

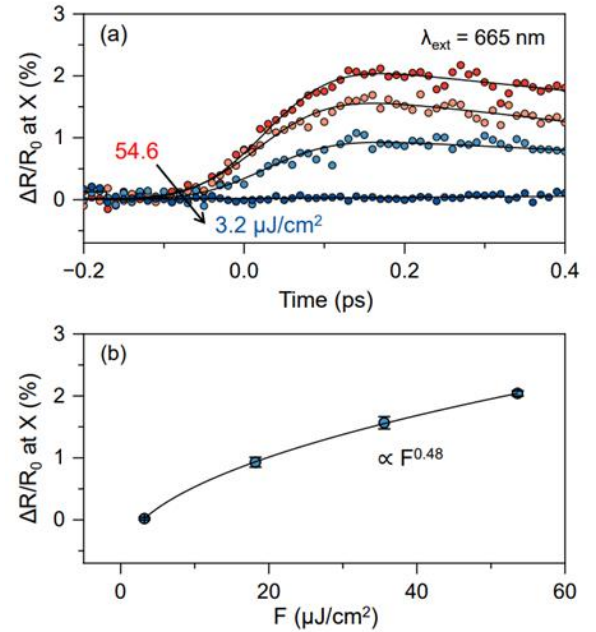

**Figure 6. Pump fluence-dependence of D → X up-conversion signal.** (a) Pump laser fluence dependent ΔR/R$_0$ signal at the X resonance (612 nm) under 665 nm excitation with fluence of 54.6, 35.6, 18.2, and 3.2 μJ/cm$^2$, respectively. (b) ΔR/R$_0$ peak intensities as a function of pump laser fluence (F) taken from (a) (circles). The solid curve presents the fit using ΔR/R$_0$ ∝ F$^k$ with k = 0.48.

→ X up-conversion as shown in Fig. 6. If TPA is responsible for the up-conversion, the X population signal induced by the D band-edge pump would increase proportionally to the square of $F_{pump}$.[68] However, we observed that the peak value of the $\Delta R/R_0$ signal at the X resonance increased with an exponent of 0.48 in relation to $F_{pump}$, which indicates a sublinear behavior typical of defect-localized states in TMDCs.[10,11,72] In contrast, we note that the $\Delta R/R_0$ signal resulting from X band-edge excitation (Fig. 5b) exhibits a linear increase over a similar $F_{pump}$ range of tens of μJ/cm² as shown in Fig. S13. Based on these observations, we ruled out the possibility of D → X up-conversion via TPA.

Phonon scattering could also account for the observed D → X up-conversion.[15] We estimated the up-conversion rate constant ($k$) by phonon scattering using the Arrhenius equation: $\boldsymbol{k = Ae^{-E_a/k_BT}}$, where $A$ is the frequency factor corresponding to different Raman mode frequencies of WS$_2$, and $\boldsymbol{E_a}$ is the activation energy for up-conversion, which is the energy difference between the X and D states and estimated to be ~ 80 meV as a lower bound, given that the true energy barrier may be larger. Using our Raman spectrum (Fig. 2b), the time constants ($1/\boldsymbol{k}$) can be estimated to be 1.7 ps for the A′$_1$, 2.1 ps for the E′$_2$, and 4.2 ps for the LA phonon modes. Additionally, as reported in previous study on multi-phonon-induced up-conversion in WS$_2$,[15] the time constants for up-conversion through multi-phonon processes were 1.5 ps for the LA+2ZA phonon mode and 1.5 ps for the 2LA+2ZA phonon mode. Thus, the expected timescale for D → X up-conversion via phonon scattering is on the order of a few picoseconds, which is considerably larger than the ~150 fs timescale that we observed.

Another possibility is coherent coupling between the X and D states. On ultrafast timescales, quantum coherence can play a significant role. The observed ultrafast (~150 fs) up-conversion process may result from coherent interactions between X and D states as seen in exciton complexes (free, charged, and bi-excitons) in TMDCs.[34,66,70,71] Quantum beating from population oscillations, a signature of the coherently coupled state,[34,66] has a coherent oscillation period of about 50 fs, given the 80 meV energy difference between X and D states. However, this period is much shorter than our temporal resolution (~100–150 fs), making it difficult to observe quantum beating directly. The spectral shape of the $\Delta R/R_0$ (Fig. 5f) exhibits the characteristics of excitation-induced dephasing with a narrow positive peak superimposed on a broad negative background.[34] Coherent population transfer occurs during the dephasing period, but in the presence of defects, the coherence time is significantly reduced and inhomogeneous broadening of the linewidth occurs.[66] This shortened coherence time of excitonic states could also reduce the ultrafast population of the X resonance through up-conversion from D states.

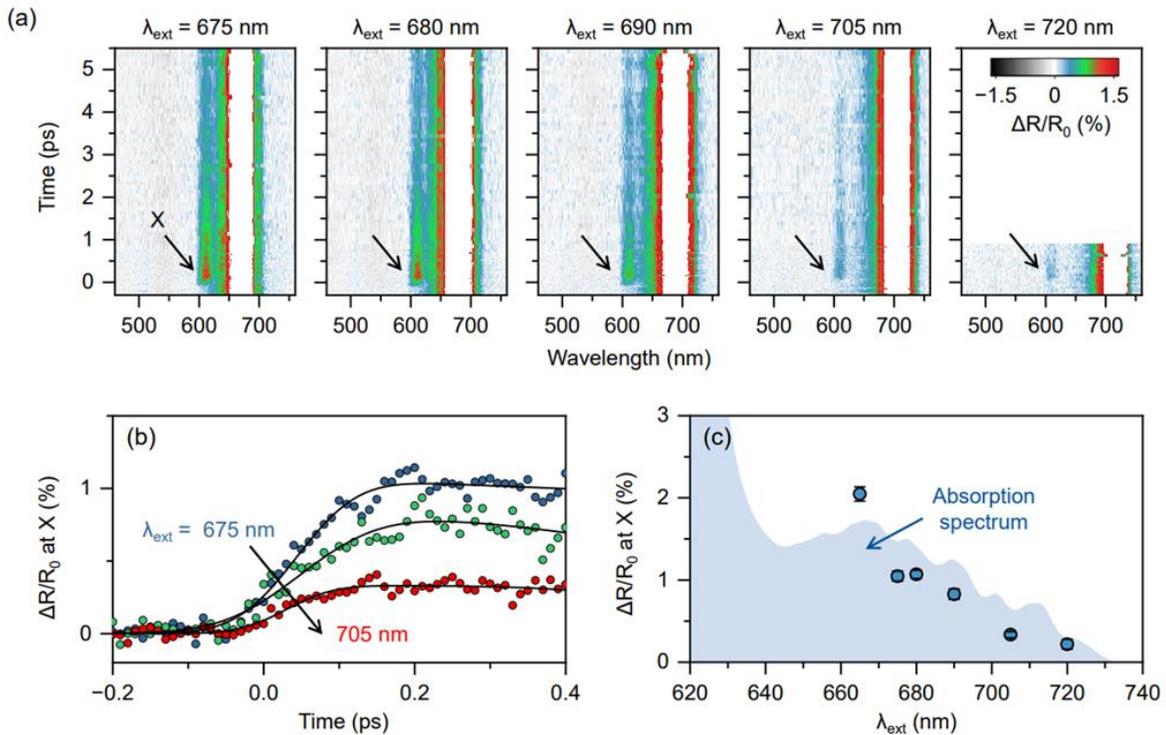

**Figure 7. Pump wavelength-dependence of D → X up-conversion signal.** (a) Excitation-wavelength ($\lambda_{ext}$)-dependent $\Delta R/R_0$ maps as a function of time and probe wavelength. $\lambda_{ext}$ indicates the pump laser wavelength. The black arrow shows the intensity of the X resonance by D → X up-conversion. (b) Time evolution of the X population upon the D band-edge excitation at 675 (blue), 690 (green), and 705 nm (red), respectively. (c) $\Delta R/R_0$ peak values as a function of excitation-wavelength taken from (a). The shaded blue area represents the part of the absorption spectrum of sample #2 measured at the center location taken from Fig. S5 with appropriate scaling.

To further understand the origin of the ultrafast up-conversion, we tuned the pump to longer wavelengths up to 720 nm – *i.e.*, a detuned photon energy ($E_X - E_{pump}$) of up to 300 meV – as shown in Fig. 7a. Remarkably, the population of the X resonance was observed even at a pump wavelength of 720 nm. It is important to note that the X-resonance population through up-conversion was not detectable in sample #1 with ~200 meV lower pump excitation (705 nm) as indicated in Fig. S14. The rise times of the $\Delta R/R_0$ signal at the X-resonance population (612 nm) across different pump wavelengths were approximately 160–170 fs, showing no systematic trend as a function of pump wavelength (Fig. 7b and Fig. S15). The small differences in rise time at different pump wavelengths are more likely attributable to experimental uncertainties, such as wavelength-dependent pulse duration or fitting error, rather than to intrinsic changes in up-conversion dynamics. As the pump wavelength increased, the X population resulting from up-conversion gradually diminished within a comparable pump laser fluence range (~30 μJ/cm$^2$). This trend aligns with the D emission spectrum shown in Fig. 7c.

Revisiting the phonon-mediated process, the activation energy for the 705 nm pump is approximately 1.9 times greater than that of the 665 nm pump. This suggests that the up-conversion timescale should be about 108 times longer at 705 nm compared to the 665 nm pump. Consequently, a significantly greater number of phonons would need to be involved in the up-conversion process, leading to a much longer observed rise time. However, the rise times for both the 705 nm and 665 nm pumps are approximately the same, around 150 fs. Therefore, the ultrafast up-conversion population observed even at a lower photon energy (with ~300 meV detuning from the X spectral position) can be attributed to coherent coupling rather than a phonon-mediated process.

Previously reported cases of trion → exciton and A exciton → B exciton up-conversions serve as good examples.[34,66] Trion → exciton up-conversion with a ~30 meV energy splitting occurs over ~8 ps,[66] well-explained by a phonon-mediated process and consistent with the few-ps timescale predicted by the Arrhenius equation above. In contrast, A → B exciton up-conversion is observed on an ultrafast timescale of tens of fs at ~380 meV energy splitting, attributed to a Dexter-like interaction[34,36] where excitons in different valleys, but with the same spin, are directly coupled.[35,36] This mechanism is explained as short-range dipole-dipole coupling between optically induced exciton polarizations in different valleys, leading to coherent population transfer.[34,35] This suggests the possibility of coherently coupled D and X states enabled by Dexter-like interactions. In this case, D → X up-conversion would occur much faster than our temporal resolution, explaining the observed population transfer time being independent of excitation wavelength. In this scenario, the observed ultrafast D → X up-conversion implies existence of a spin-valley polarization in the defect states, which has been theoretically predicted[31,32] and experimentally revealed by Zeeman splitting in defect-bound exciton emission.[30] These coherently coupled states allow instantaneous up-conversion, supporting the observation of population transfer even with photoexcitation significantly below the X resonance (~300 meV).

**CONCLUSION**

Using alkali metal halide-assisted CVD method, we synthesized monolayer WS$_2$ crystals that exhibit strong defect-induced PL at room temperature with an energy ~80 meV lower than that of free excitons. By comparing PL and

Raman maps, we found that the intensity ratio of LA to $A'_1$ phonon modes closely correlates with this low-energy PL emission, indicating the presence of defect-bound excitons. Atomic-resolution STEM imaging revealed two dominant defect types – mono-sulfur vacancies ($V_S$) and W-site defect complexes ($S_WV_S$) – with $V_S$ densities approximately twice as high at the center of the monolayer crystal compared to the edges. This spatial variation, coupled with minimal variation in $S_WV_S$ defects, suggests that the observed defect emission band primarily originates from $V_S$ defects. KPFM further confirmed higher defect densities in the center of the $WS_2$ crystal from sample #2, evidenced by a lower local work function.

Ultrafast optical spectroscopy revealed new ultrafast dynamics associated with these defect states. By tuning the pump wavelengths and fluences, we directly observed ultrafast defect-bound exciton dynamics. Under photoexcitation with a 400 nm pump above band-edge, both free and defect-bound excitons were formed nearly simultaneously (~300 fs) following hot carrier relaxation, significantly faster than previously reported broad timescales of 1–100 ps.[20-26] However, the defect-bound exciton lifetime was shorter than that of the free exciton, leading to a population difference on a 1–100 ps timescale, which could further drive energy-favored incoherent exciton trapping. This explains the trapping timescales reported in prior studies,[13,20-26] which are estimated based on the reduction in carrier and exciton lifetimes by defects.

Under band-edge photoexcitation, we observed unexpected fast population interconversion (~150 fs) between free and defect-bound excitons, a timescale comparable to our instrument temporal resolution. We ruled out two-photon absorption (TPA) as the cause of the up-conversion (defect-bound exciton to free exciton) through pump fluence-dependent experiments. Additionally, the up-conversion process remained ultrafast (~150 fs) even when using longer pump wavelengths (~100 nm longer corresponding to ~300 meV lower energy relative to the X band). This suggests that phonon-mediated processes are unlikely, as they typically occur on longer timescales. Instead, we attribute this rapid interconversion from D to X states to coherent coupling between free and defect-bound excitons. The estimated oscillation period (~50 fs) of quantum beating between the X and D states is shorter than the temporal resolution of our setup. Direct observation of coherent X–D beating would require sub-10-fs, phase-stable laser pulses. Nevertheless, alternative mechanisms such as bandgap renormalization, two-photon absorption, and multi-phonon coupling can be excluded based on spectral-, wavelength-, and fluence-dependent analyses. The observed ultrafast interconversion between X and D excitons thus provides a basis for future time-domain studies using multidimensional or sub-10 fs spectroscopy. These findings reveal that population transfer between X and D states in defect-engineered $WS_2$ can occur on sub-150 fs timescale (limited by our instrument resolution) – orders of magnitude faster than previously reported. Beyond advancing fundamental understanding of defect-mediated exciton ultrafast dynamics in TMDCs, our results offer valuable guidance for defect-engineering in 2D materials for optoelectronic applications where defect-bound excitons could facilitate efficient energy conversion through coherent up-conversion processes.

## METHODS

**Sample preparation.** Samples #1 and #2 of monolayer $WS_2$ were synthesized by low pressure and atmospheric pressure CVD on $Si/SiO_2$ substrates, respectively. For sample #1, a 30 nm-thick $WO_3$ film was deposited onto a $Si/SiO_2$ substrate (285nm-thick thermal $SiO_2$ layer) by electron beam evaporation from a $WO_3$ pellet (Kurt J. Lesker, 99.99%), and was then covered with another clean $Si/SiO_2$ substrate, creating a sandwich with $SiO_2$ surfaces facing each other in direct contact. This assembly was placed at the center of a 2″ quartz tube furnace, while 150mg sulfur powder (Alfa Aesar, 99.5%) was positioned upstream outside the central heating zone. The chamber was evacuated to a base pressure of 10 mTorr using a mechanical pump. The growth was performed at 100 Torr of ultrahigh purity argon flowing at 100 sccm (standard cubic centimeters per minute) that was introduced starting at 150 °C to minimize moisture and residual gases. The heating rate was 15 °C/min and the growth was conducted at 825 °C for 15 min. Under optimized location of the sulfur crucible in the quartz tube outside the high temperature zone, sulfur evaporation began when the furnace central zone reached 750 °C. After the growth was finished, the furnace was naturally cooled down to 600 °C within 10 minutes and was then opened to allow more rapid cooling. More details on the sample growth using this approach can be found elsewhere.[73]

For sample #2, Si/SiO$_2$ substrates were placed with SiO$_2$ surface facing down above an alumina crucible containing ~2 mg powder of WO$_3$ (Sigma-Aldrich, 99.995% metal basis) mixed with ~1 mg NaBr powder (Sigma-Aldrich, 99.99% metal basis) in 1:0.5 weight ratio. This mixture was designated as the precursor. The crucible with the WO$_3$ and NaBr mixture precursor and substrate was loaded at the center of a 1"-diameter quartz tube in a tube furnace. Another crucible containing ~200 mg sulfur powder was located in the upstream side of the tube outside the furnace where heating tape was wrapped. After flushing the tube with 500 sccm of ultrahigh purity argon gas, the atmospheric pressure argon gas flow was reduced to 50 sccm, and the precursor was heated to 780 °C at a ramping rate of 40 °C/min, after which the sulfur was heated to ~250 °C by the heating tape. The sulfur evaporation was performed for 3 minutes, after which the heating tape was removed, and the furnace was opened to allow rapid fan-assisted cooling to room temperature.

**PL and Raman mapping.** The PL spectra were measured in a custom-built micro-PL setup. The PL was excited with a continuous wave laser at 532 nm through an upright microscope using a 100× objective with NA (numeric aperture) = 0.9 (beam spot ~1 μm). The typical incident laser power on a sample was maintained at approximately 3.6 μW to reduce possible laser heating and sample damage during PL spectra acquisition. The PL light was analyzed by a spectrometer (Spectra Pro 2300i, Acton, $f$ = 0.3 m) that was coupled to the microscope and equipped with 150, 600, and 1800 grooves/mm gratings and a CCD camera (Pixis 256BR, Princeton Instruments). An 1800 grooves/mm grating was used for Raman measurements with an incident laser power of 270 μW. The PL and Raman mapping were conducted using a motorized two-axis stage.

**Micro-absorption measurements.** The absorption spectra are measured using a laser-driven white light source (EQ-99FC, Energetiq). The white light source is coupled out to a 25 μm diameter optical fiber and is focused on a sample surface to a ~ 1.5 μm spot using two microscope objectives: a 5x collimating objective, NA (numeric aperture) =0.1 and a 100× long working distance objective, NA=0.8 in an inverted microscope. The transmitted light is collected by a 50x long working distance objective (NA=0.5) in an upright microscope coupled to the inverted microscope and is analyzed by a spectrometer (Spectra Pro 2300i, Acton) equipped with a CCD camera (Pixis 256BR, Princeton Instruments). The absorbance (A) was calculated as A = log$_{10}$(I$_0$/I), where I and I$_0$ are the light intensities transmitted through the substrate on and off a TMD crystal, respectively.

**Up-conversion PL measurements.** These measurements were performed using a triple spectrometer (Jobin-Yvon T64000) equipped with three 1800 grooves/mm gratings and a CCD detector (Synapse Plus, Horiba) under a microscope in a backscattering configuration. A tunable (450-525 nm and 540-650 nm) visible CW optical parametric oscillator (OPO) (C-Wave, Hubner Photonics) was used as PL excitation source. Excitation laser light was focused onto the sample with a 100× objective (numeric aperture N/A = 0.9) to a spot size of about 1 μm. The excitation laser power on the sample was varied from 0.3 mW to 7 mW.

**Kelvin probe force microscopy.** Kelvin Probe Force Microscopy (KPFM) measurements were conducted using an Asylum Research Cypher (an Oxford Instruments Company) microscope in single-pass amplitude modulation (AM-KPFM) mode. The sample was mounted onto a conductive sample puck using silver paint, ensuring good electrical contact and grounding with respect to the device. A Pt-coated Multi75G probe (nominal resonance frequency ~75 kHz, spring constant ~3 N/m, tip radius <25 nm) was used for both topography and surface potential measurements. KPFM was operated well below resonance at 5 kHz with an applied AC voltage of 3 V to detect the electrostatic force component directly, enabling surface potential mapping during the same pass as topography acquisition. This setup allowed for simultaneous capture of morphological and work function variations across the sample surface.

**Electron microscopy.** $WS_2$ monolayer crystals from both samples were transferred on TEM grids and were cleaned prior to imaging via radiolysis of adsorbed water and electron beam shower treatment as described elsewhere.[74] HAADF-STEM images were taken on a probe corrected Thermo Scientific Spectra 300, operating at 60 kV accelerating voltage with a convergence angle of 30 mrad and a nominal screen current of 50 pA. The HAADF collection angle was 51-200 mrad. Simulated HAADF-STEM images were generated with the ASE and abTEM Python packages, under the same conditions with the exception of beam current – the low dose images are simulated using $10^4$ electrons and the high dose images are simulated with $10^{10}$ electrons. HAADF-STEM atom finding routines involve a Lucy-Richardson deconvolution, blob finding, and refinement of the blobs into atom positions via 2D Gaussian fitting.[75] Identification of the W-complexes is done through a breadth first search, where each tungsten vacancy is associated with 6 coordinates, one for every neighboring tungsten atom. The positions of the tungsten vacancies are then obtained with a density-based spatial clustering of applications with noise (DBSCAN). These routines are implemented in Python package pyTEMlib.

**Ultrafast transient absorption spectroscopy.** Linear reflectance spectra were measured at the same sample location using a laser driven broadband light source (Energetiq Inc.). Transient reflectance spectroscopy was performed using a Ti:Sapphire oscillator (Micra, Coherent) with its output used to seed a Ti:Sapphire Coherent Legend (USP-HE) amplifier operating at 1 kHz repetition rate. The Legend amplifier provides pulses centered at 800 nm, with ~57 fs duration and 1.8 mJ pulse energy. A white-light continuum probe pulse was generated by focusing a small portion of the beam from the Legend amplifier onto a sapphire window (2-mm thick). Pump pulses (3.1 eV) for above band-edge photoexcitation are obtained by second harmonic generation of the fundamental output of the Legend amplifier. For band-edge photoexcitation, we used an optical parametric amplifier (TOPAS, Coherent) to tune the pump wavelengths from 600 to 720 nm with ~100 fs duration. Collinear pump and probe beams were introduced into a custom-built upright microscope equipped with a 36× reflective objective resulting in the pump and probe spot diameters ~5 μm at the sample location. The pump power was adjusted with neutral density filters. A spectrograph (Shamrock 303i, Andor) coupled to an EMCCD (Andor Newton) was used for detecting the reflected white light probe. The pump and probe pulses were cross-polarized, and we used a polarizer before the spectrometer to reduce the pump scattering signal in $\Delta R/R_0$ map. At each time delay, $\Delta t$, between the pump and the probe, the change in the reflectance $\Delta R/R_0$ was calculated as $(R_{pump} - R_{no\ pump})/R_{no\ pump}$, where $R_{pump}$ and $R_{no\ pump}$ are the intensities of the reflected probe with pump on and off, respectively. We further performed chirp correction of $\Delta R/R_0$ map based on a cross-correlation function measurement below.

**Cross-correlation function measurements.** To characterize the temporal chirp and resolution of our ultrafast pump-probe setup, we conducted cross-correlation function measurements using 1.0 mm thick fused silica substrate. Figure S10 presents a 2D plot of $\Delta R/R_0$ as functions of time delay and probe wavelength, reflecting the measurement of supercontinuum chirp with 400 nm pump. As depicted in Fig. S10a, the temporal overlap of pump and probe changes with probe wavelength, showing approximately 1.5 ps dispersion between 450 and 750 nm, with shorter probe wavelengths arriving earlier due to a shorter optical path. The optimal temporal overlap, defined by the probe wavelength-dependent time-delay zero ($t_0$), is necessary for accurate chirp correction.

The instantaneous coherent electronic signal in transparent media, such as fused silica,[76] introduces a coherent artifact in transient measurements as shown in Fig. S10b. For linearly chirped probes, the coherent $\Delta R/R_0$ signal can be expressed by a superposition of a cross-correlation function ($F_{cc}$), and its first and second derivatives with phase term as[76]

$$\frac{\Delta R}{R_0}(\lambda, t) = \cos(t - t_0 + \varphi) \cdot (a_0 F_{cc} + a_1 \frac{\partial F_{cc}}{\partial t} + a_2 \frac{\partial^2 F_{cc}}{\partial t^2}) \quad \text{(Eq. 1)}$$

where, $\varphi$ is the phase of observed coherent oscillation signal at probe wavelength. We assumed the $F_{cc}$ of electric field of pump and probe pulses has a Gaussian temporal response with temporal width ($\tau$) as $F_{cc} = e^{-2(t-t_0)^2/\tau^2}$. By fitting the coherent $\Delta R/R_0$ signal with Equation 1 (Fig. S10b), we obtained the $t_0$ as a function of probe wavelength as shown in Fig. S10c. The obtained probe-wavelength dependent $t_0$ is fitted with a second order polynomial function as $t_0(\lambda) = -11.75 + 0.038\lambda - 2.67 \times 10^{-5}\lambda^2$. This obtained dispersion of was used for chirp correction of measured $\Delta R/R_0$ map.

With fitting results, we also estimated the temporal resolution of our setup. The temporal resolution is usually defined with the full width at half maximum (FWHM) based on the measured $\tau$ as FWHM = $(2ln2)^{0.5}\tau$. The fitting error of $\tau$ was less than ±10 fs. As shown in Fig. S10d, the FWHM temporal resolution is slightly depending on the probe wavelength and exhibits ~100–150 fs. In terms of temporal resolution as 90% to 10% of its full value (the 90%-10% edge width) can be calculated by 1.28$\tau$, resulting in a slightly larger value of ~130 fs in the 550–700 nm probe wavelength range. These values are consistent with the observed rise time of interconversion between free and defect-bound excitons. The temporal resolution of the cross-correlation function with a 600 nm pump using OPA shows similar values, as shown in Fig. S10d, corroborated by measurements with an OPA pump and white-continuum probe (Fig. S10e).

## ASSOCIATED CONTENT

**Supporting Information**.

The Supporting Information is available free of charge at https://...

> Distributions of Raman peak and intensity ratios (Figure S1), surface potential mapping via KPFM (Figure S2), atomic-resolution STEM images and defect identifications in WS$_2$ monolayers with and without NaBr (Figures S3–S4), steady-state reflectance/absorption spectra (Figure S5), PL, Raman, absorption characterization of a WS$_2$ monolayer crystal synthesized on a quartz substrate with the addition of the alkali-metal halide NaBr (Figure S6), transient reflectance spectra under above band-edge excitations (Figure S7), pump spectral characteristics for excitation-tuned measurements (Figure S8), transient reflectance spectra under X band-edge excitations (Figure S9) cross-correlation data and temporal resolution analysis (Figure S10), up-conversion PL measurements (Figure S11), transient reflectance spectra around X band under D band-edge excitations (Figure S12), fluence-dependent transient dynamics and signal amplitudes in X pump-X probe (Figure S13), $\Delta R/R_0$ map of WS$_2$ grown without alkali-metal halide NaBr sample #1 at 705 nm pump wavelength (Figure S14), pump wavelength-resolved temporal evolution of transient reflectance at X resonance (Figure S15).

## AUTHOR INFORMATION

**Corresponding Authors**

**Kai Xiao** – Center for Nanophase Materials Sciences, Oak Ridge National Laboratory, Oak Ridge, Tennessee 37831, Email: xiaok@ornl.gov

**Alexander A. Puretzky** – Center for Nanophase Materials Sciences, Oak Ridge National Laboratory, Oak Ridge, Tennessee 37831, Email: puretzkya@ornl.gov

**Authors**

**Tae Gwan Park** – Center for Nanophase Materials Sciences, Oak Ridge National Laboratory, Oak Ridge,

Tennessee 37831, Center for Integrated Nanotechnologies, Los Alamos National Laboratory, Los Alamos, New Mexico 87544

**Xufan Li** – Honda Research Institute USA Inc., San Jose, California 95134

**Kyungnam Kang** – Center for Nanophase Materials Sciences, Oak Ridge National Laboratory, Oak Ridge, Tennessee 37831

**Austin Houston** – Department of Materials Science and Engineering, University of Tennessee, Knoxville, Tennessee 37996

**Liam Collins** – Center for Nanophase Materials Sciences, Oak Ridge National Laboratory, Oak Ridge, Tennessee 37831

**Gerd Duscher** – Department of Materials Science and Engineering, University of Tennessee, Knoxville, Tennessee 37996

**David B. Geohegan** – Center for Nanophase Materials Sciences, Oak Ridge National Laboratory, Oak Ridge, Tennessee 37831

**Christopher M. Rouleau** – Center for Nanophase Materials Sciences, Oak Ridge National Laboratory, Oak Ridge, Tennessee 37831

**Author Contributions**

All authors have given approval to the final version of the manuscript.

**Notes**

The authors declare no competing financial interest.

**ACKNOWLEDGMENT**

This work was supported by the Center for Nanophase Materials Sciences (CNMS), which is a US Department of Energy, Office of Science User Facility at Oak Ridge National Laboratory. Part of synthesis and STEM characterization were supported by the U.S. Department of Energy, Office of Science, Basic Energy Sciences, Materials Sciences and Engineering Division. The scanning transmission electron microscopy work was performed at the Institute for Advanced Materials and Manufacturing (IAMM) at the University of Tennessee, Knoxville. Partial support for this work was provided by the Los Alamos National Laboratory Laboratory Directed Research and Development (LDRD) programme (Project number: 20230124ER). Los Alamos National Laboratory, an affirmative action equal opportunity employer, is managed by Triad National Security, LLC for the US DOE NNSA, under contract no. 89233218CNA000001.

**Supporting Information for**

# Direct observation of ultrafast defect-bound and free exciton dynamics in defect-engineered WS$_2$ monolayers

Tae Gwan Park[1,2], Xufan Li[3], Kyungnam Kang[1], Austin Houston[4], Liam Collins[1], Gerd Duscher[4], David B. Geohegan[4], Christopher M. Rouleau[1], Kai Xiao[1,*], Alexander A. Puretzky[1,*]

[1]Center for Nanophase Materials Sciences, Oak Ridge National Laboratory, Oak Ridge, Tennessee 37831
[2]Center for Integrated Nanotechnologies, Los Alamos National Laboratory, Los Alamos, New Mexico 87544
[3]Honda Research Institute USA Inc., San Jose, California 95134
[4]Department of Materials Science and Engineering, University of Tennessee, Knoxville, Tennessee 37996

*Corresponding author(s) email:  xiaok@ornl.gov and puretzkya@ornl.gov

**Supplementary Figure S1**

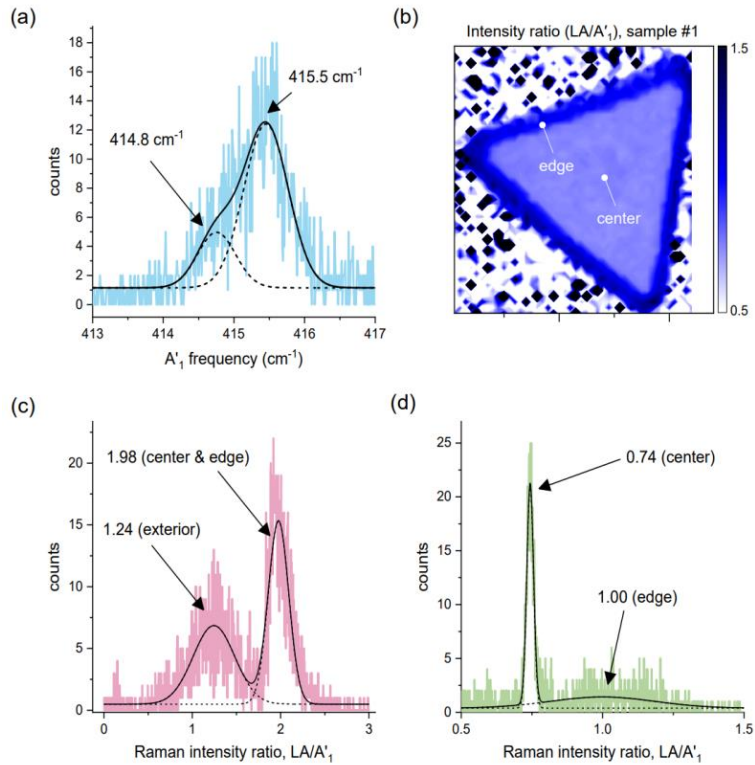

Figure S1. (a) Distribution of the $A'_1$ Raman band center obtained from the map shown in Fig. 2c of the main text. (b) Map of Raman intensity ratio, $LA/A'_1$, measured for sample #1 for comparison with the corresponding map for sample #2 presented in Fig. 2f of the main text. (c,d) Distributions of the $LA/A'_1$ intensity ratios in sample #2 (c) and #1 (d) obtained from the corresponding maps (Fig. 2f and S1b, respectively).

**Supplementary Figure S2**

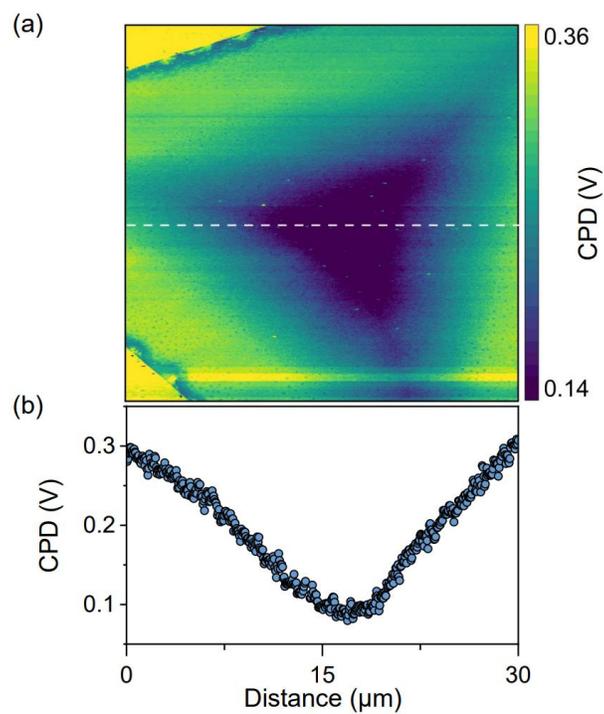

Figure S2. (a) Kelvin probe force microscopy (KPFM) image of WS$_2$ monolayer synthesized with NaBr surfactant (sample #2). (b) Corresponding contact potential difference (CPD) representing the work function difference between a tip and a sample, measured along the dashed line in (a).

**Supplementary Figure S3**

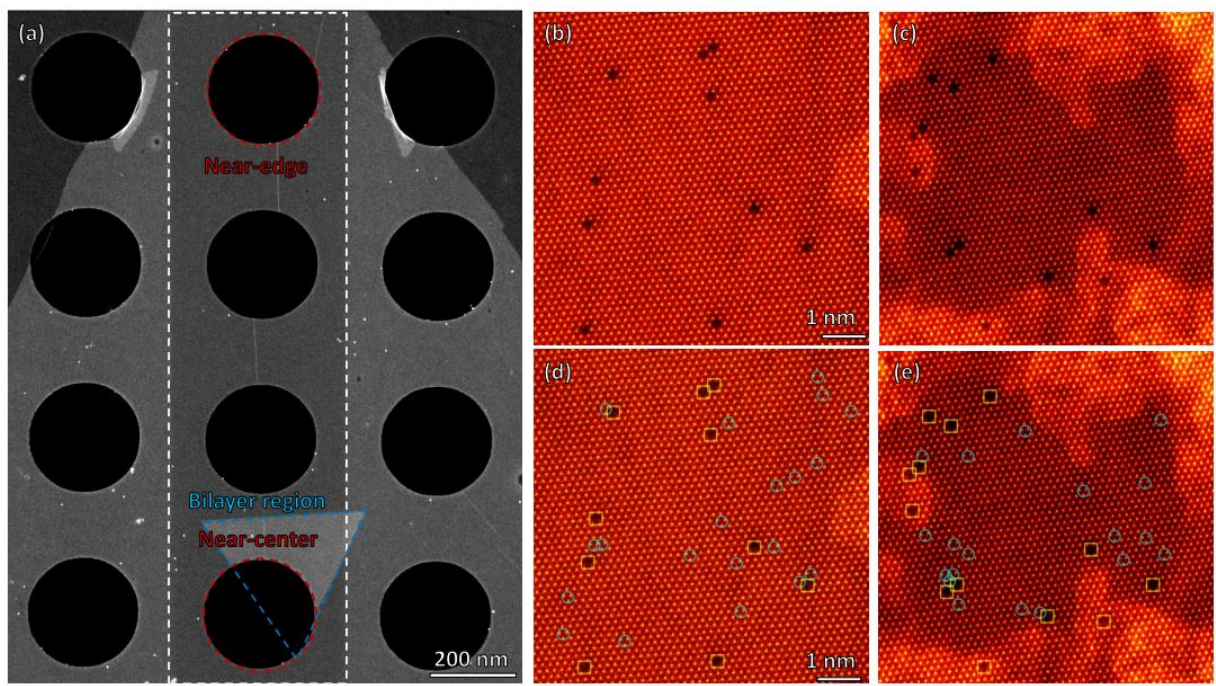

Figure S3. (a) Low-magnification image of a transferred NaBr-assisted grown $WS_2$ monolayer on a Quantifoil TEM grid that was used for the defect analysis in Fig. 4 in the main text. The red dashed circles indicate near-edge and near-center regions. The bilayer region is highlighted to point out that this is the center of the flake. The bilayer region was not used for defect analysis. (b,d) High-magnification HAADF-STEM image with defects marked from the center region in (a). (c,e) High-magnification HAADF-STEM image with marked defects from the near-edge region in (a). Monosulfur vacancies ($V_S$) and W-site defect complexes ($S_WV_S$) are marked by blue circles and yellow squares, respectively.

**Supplementary Figure S4**

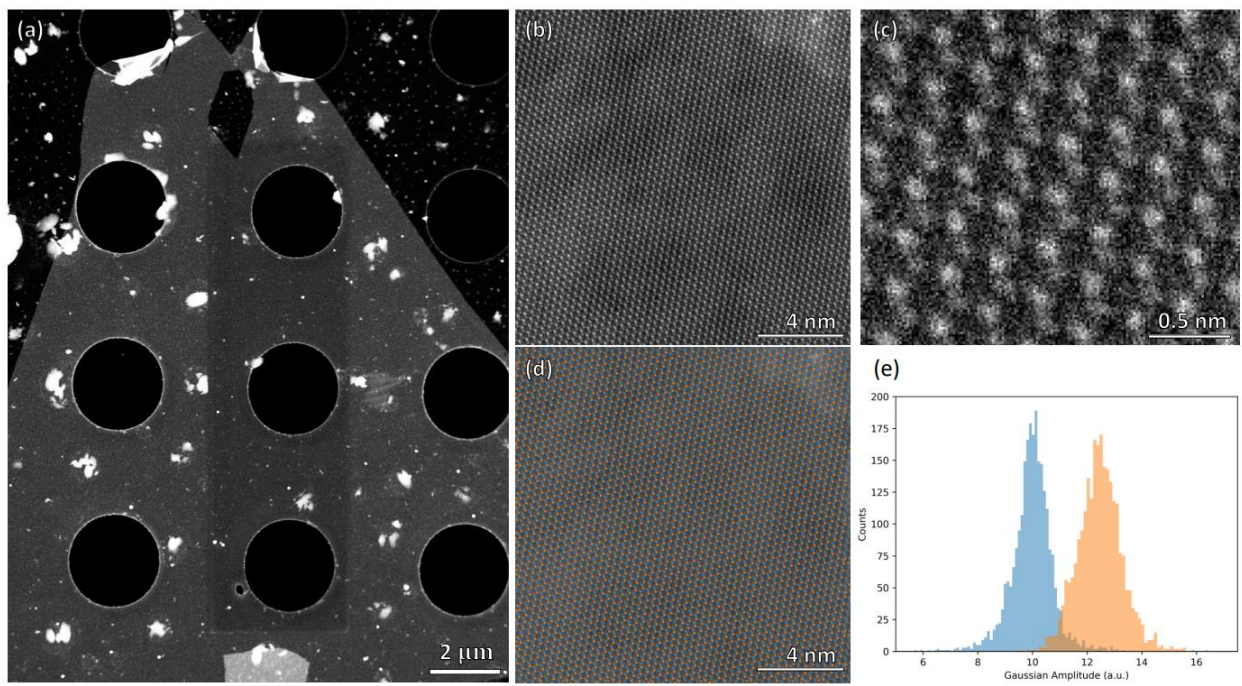

Figure S4. (a) Low-magnification image of a transferred WS$_2$ monolayer grown without NaBr on a Quantifoil TEM grid. (b) HAADF-STEM image of the WS$_2$ monolayer from (a). There are no discernible S vacancies. (c) A magnified selection from (b) showcasing the resolution and image quality used for analysis. (d) The same image as (b) with W atoms and S atomic stacks marked in blue and orange, respectively. (e) The intensities of the Gaussian fit to each atom/atomic stack. Notably, the S-intensity distribution shows no discernible left shoulder where the intensity of S vacancies would be expected if there were any in the image.

**Supplementary Figure S5**

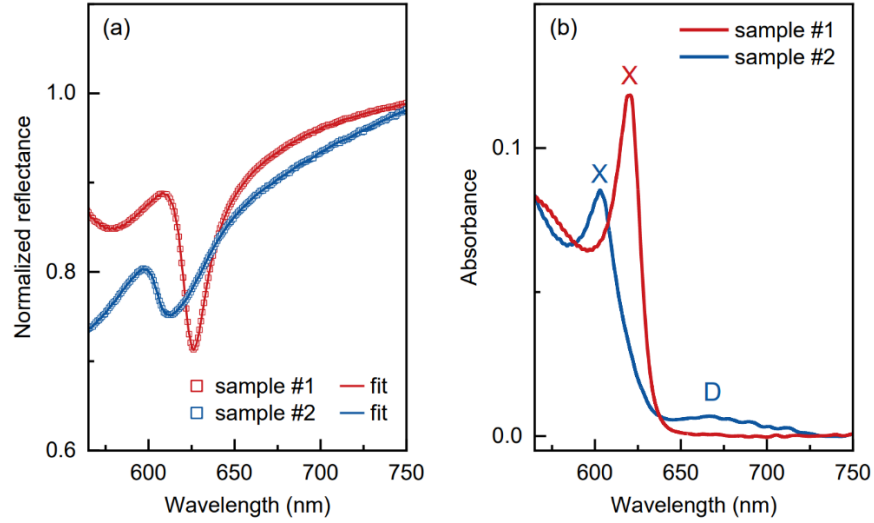

Figure S5. (a) Normalized reflectance ($R_0 = R_{sam}/R_{sub}$) and fitting results based on Kramers–Kronig constrained variational analysis for samples #1 and #2. (b) Absorption spectra of WS$_2$ monolayers obtained from samples #1 and #2.

We resolved a broad D exciton feature on the low-energy side of X in the normalized reflectance spectrum as shown in Fig. S5a. However, optical interference from multiple reflections on SiO$_2$/Si substrates limits a clean separation of the X and D absorption bands. To mitigate this, we convert the normalized reflectance $R_0 = R_{sam}/R_{sub}$ into an effective absorbance by modeling the multilayer system consisting of the WS$_2$/SiO$_2$/Si and performing a Kramers–Kronig (KK) constrained variational analysis.[1] For normal incidence, the sample and substrate reflectance are written:[2,3] $R_{sam} = \left| -\frac{Cn_{SiO_2} - D(1-A)}{Cn_{SiO_2} + D(1-A)} \right|^2$, $R_{sub} = \left| -\frac{Cn_{SiO_2} - D}{Cn_{SiO_2} + D} \right|^2$, where $A(\omega) = -\frac{i\omega d_{WS_2}}{c}\tilde{\varepsilon}(\omega)$ encodes the WS$_2$ monolayer response ($d_{WS_2} = 0.67$ nm), and c is the speed of The real part of $A$ determines the absorption spectrum of the WS$_2$ monolayer. C and D are functions of the refractive indices ($n_{SiO_2}$, $n_{Si}$) and phase shifts ($\delta_{SiO_2} = n_{SiO_2}\omega d_{SiO_2}/c$ and $\delta_{Si} = n_{Si}\omega d_{Si}/c$), with thicknesses of SiO$_2$ (285 nm) and Si (500 μm) as: $\begin{pmatrix} C \\ D \end{pmatrix} = \begin{pmatrix} n_{SiO_2}\sin\delta_{SiO_2} & in_{Si}\cos\delta_{Si} \\ in_{SiO_2}\cos\delta_{SiO_2} & n_{Si}\sin\delta_{Si} \end{pmatrix} \begin{pmatrix} \sin\delta_{Si} + in_{Si}\cos\delta_{Si} \\ i\cos\delta_{Si} + n_{Si}\sin\delta_{Si} \end{pmatrix}$. Optical constants of SiO$_2$ and Si are taken from the literature.[4,5] We implement a KK-constrained variational analysis by representing the WS$_2$ dielectric function as $\tilde{\varepsilon}(\omega) = \varepsilon_{cont}(\omega) + \sum_{n=1}^{N} \frac{\omega_{p,n}^2}{\omega_{0,n}^2 - \omega^2 - i\gamma\omega}$, using $N = 100$ equally spaced oscillators over the spectral range and a fixed linewidth $\gamma = 25$ meV. The oscillator strengths $\{\omega_{p,n}^2\}$ are fitting parameters that yield a smooth description of the spectra in Fig. S5a consistent with the KK relation, from which we extract the absorbance spectrum of the WS$_2$ monolayer as shown in Fig. S5b. In the absorbance spectrum, a broad feature on the low-energy side of the X band is observed only in sample #2 (defect-engineered, alkali-assisted CVD-grown WS$_2$ monolayer). In the pump–probe measurements, static substrate multiple-reflection contributions cancel in $\Delta R$ (pump on– pump off). Consequently, the X (around 610 nm) and D (around 660 nm) features appear as positive peaks in the $\Delta R/R_0$ signal, corresponding to absorption bleaching at the X and D resonances, respectively.

**Supplementary Figure S6**

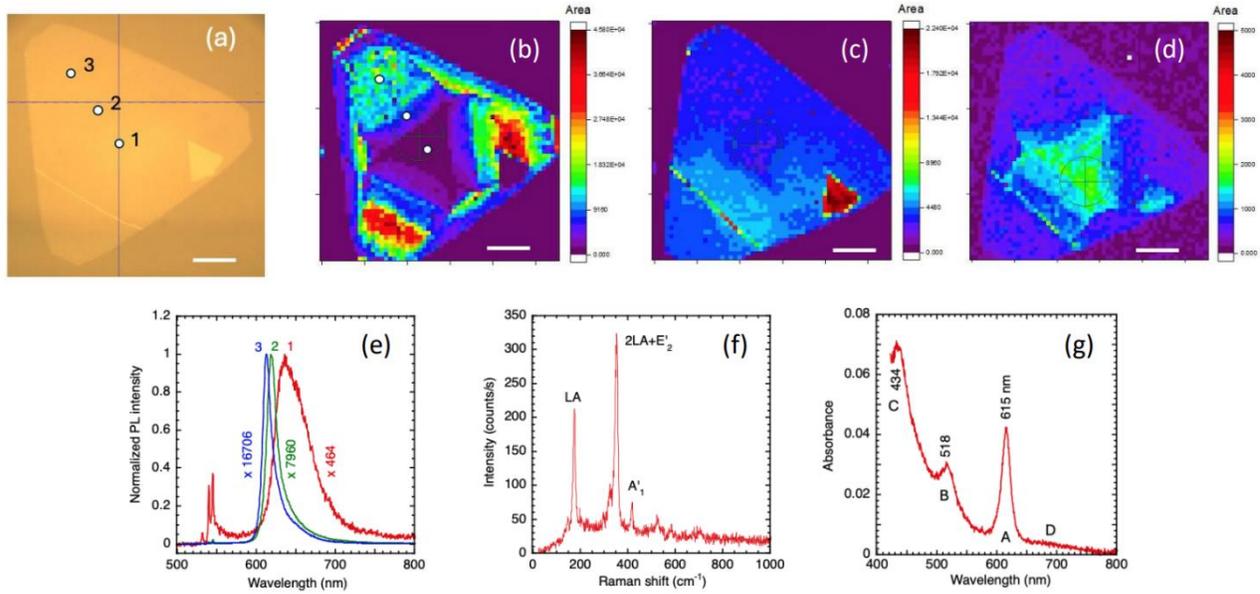

Figure S6. PL, Raman, and absorption characterization of a $WS_2$ monolayer crystal synthesized on a quartz substrate with the addition of the alkali-metal halide NaBr. (a) Optical microscope image. (b) PL map integrated over PL spectra at each point from 600 nm to 700 nm. (c,d) Raman maps obtained by integrating the $2LA+E'_2$ peak from 330 cm$^{-1}$ to 370 cm$^{-1}$ and the LA peak from 160 cm$^{-1}$ to 190 cm$^{-1}$, respectively. (e) PL spectra taken from the points 1–3 marked by the white circles in (a) and (b). The PL spectra are normalized to their maximum intensity values listed in (e) for each spectrum (in counts/s). (f) Raman spectrum measured at the point marked with a cross in (d), showing a strong LA mode related to abundant defects in the $WS_2$ monolayer crystal. The scale bars in (a–d) are 10 μm. (g) Absorption spectrum measured near the center position.

We further conducted direct optical absorption measurements using a transparent substrate. To perform these measurements, we synthesized $WS_2$ flakes on a quartz substrate by a NaBr-assisted CVD approach (same as sample #2). We first characterized it again using PL and Raman spectroscopy as shown in Figs. S6(a–f). As discussed in the main text, defect-bound exciton PL and a highly defective area were consistently observed near the center of the NaBr-assisted $WS_2$ flake. Interestingly, the PL map shows a triangular dark region that is rotated by 60° compared to that in sample #2 (Fig. 1e in the main text), with similar but slightly blue-shifted PL peaks as shown in Fig. S6e. The linear absorption spectrum in Fig. S6g clearly shows the X resonance at 615 nm. We can also observe that 8 the defect-related band, D, is much broader and substantially less intense compared to the X resonance and is similar to that obtained by the Kramers–Kronig constrained variational analysis in Fig. S5.

**Supplementary Figure S7**

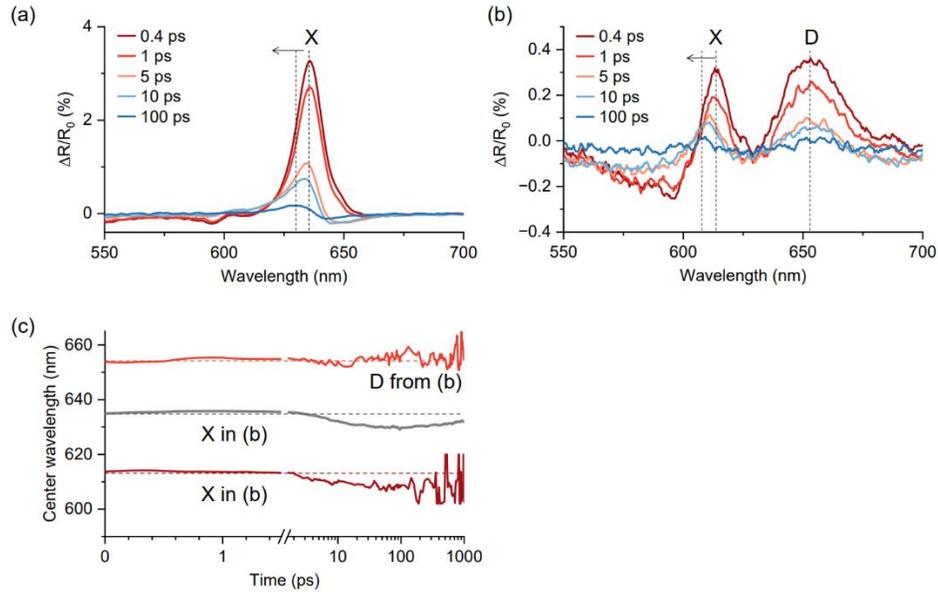

Figure S7. (a,b) ΔR/R$_0$ spectra at selected time delays with above band-edge photoexcitation (400 nm) in samples #1 and #2, respectively. (c) Time-dependent peak wavelength of ΔR/R$_0$ spectra at the X and D resonances from (a) and (b), respectively. The center wavelengths at each time delay were obtained by fitting a local window around the X and D resonances with a Gaussian peak function, yielding the center wavelength and amplitude.

**Supplementary Figure S8**

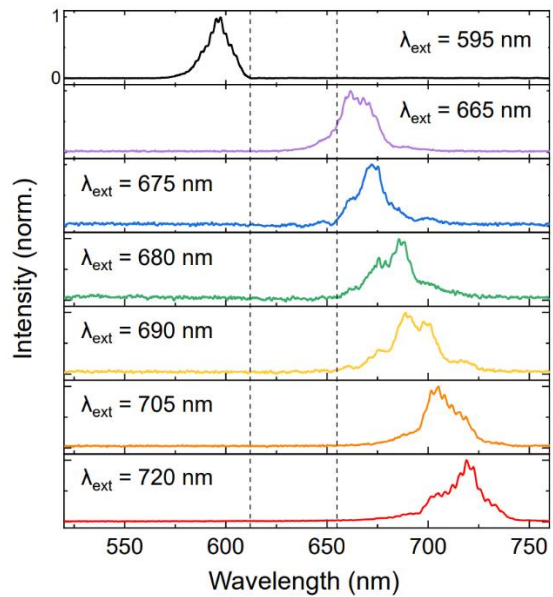

Figure S8. Pump spectrum for excitation-tuned transient absorption spectroscopy presented in Figs. 5 and 7 in the main text. Two vertical dashed lines at 612 and 655 nm indicate the positions of the X and D bands, respectively. For each X and D band-edge excitation experiment, we ensured that no other bands were directly excited.

**Supplementary Figure S9**

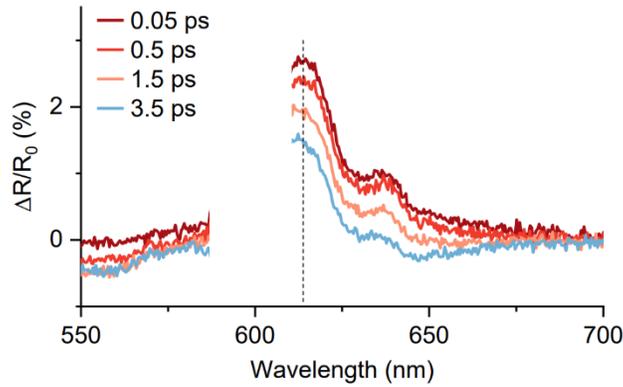

Figure S9. $\Delta R/R_0$ spectra at selected time delays with X band-edge photoexcitation (595 nm) in sample #2.

**Supplementary Figure S10**

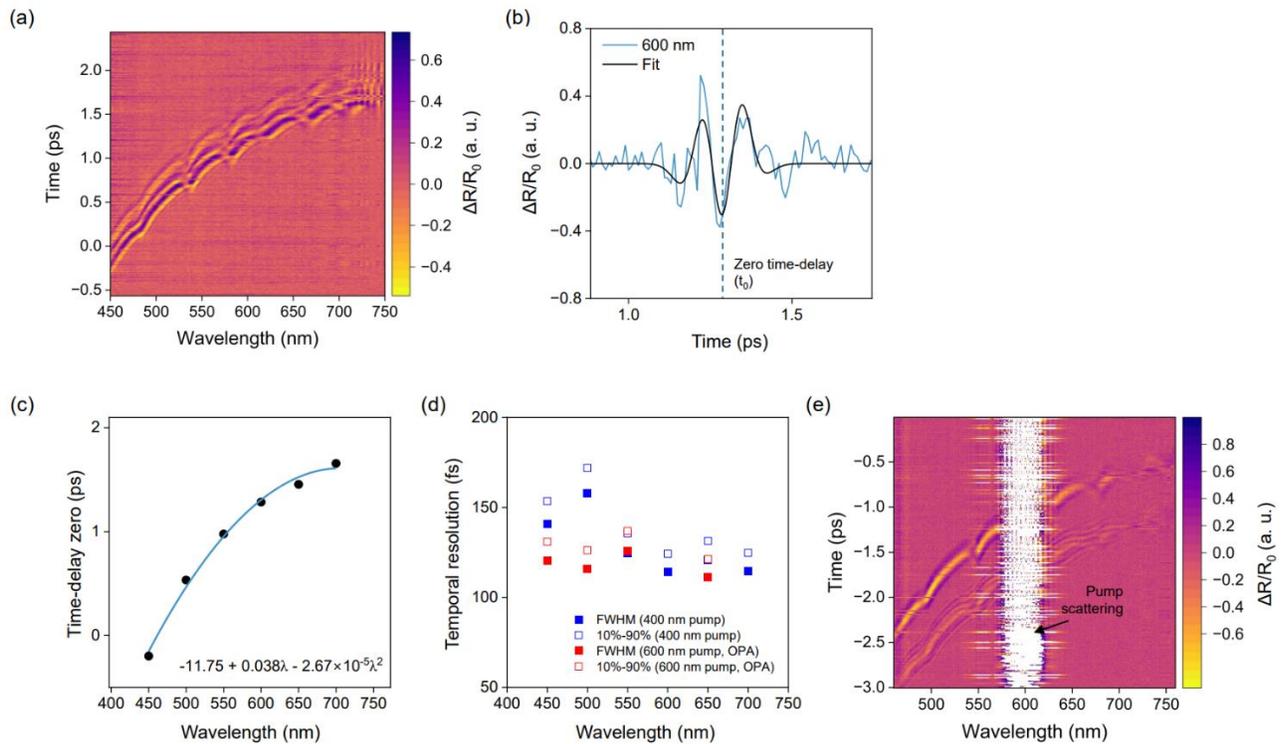

Figure S10. (a) Cross-correlation measurement at 400 nm pump and white-continuum probe with a 1.0 mm-thick fused silica substrate. (b) Cross-correlation signal at a 600 nm probe wavelength. The black curve indicates the fitting result based on a superposition of a cross-correlation function between the pump and the supercontinuum probe, and its first and second derivatives, as described in the Methods section. The same fitting at different probe wavelengths gives the (c) time-delay zero, (d) full-width at half maximum, and (e) 10%–90% temporal resolution as functions of probe wavelength, as well as the observed values in the same measurement with a 600 nm pump using the OPA.

**Supplementary Figure S11**

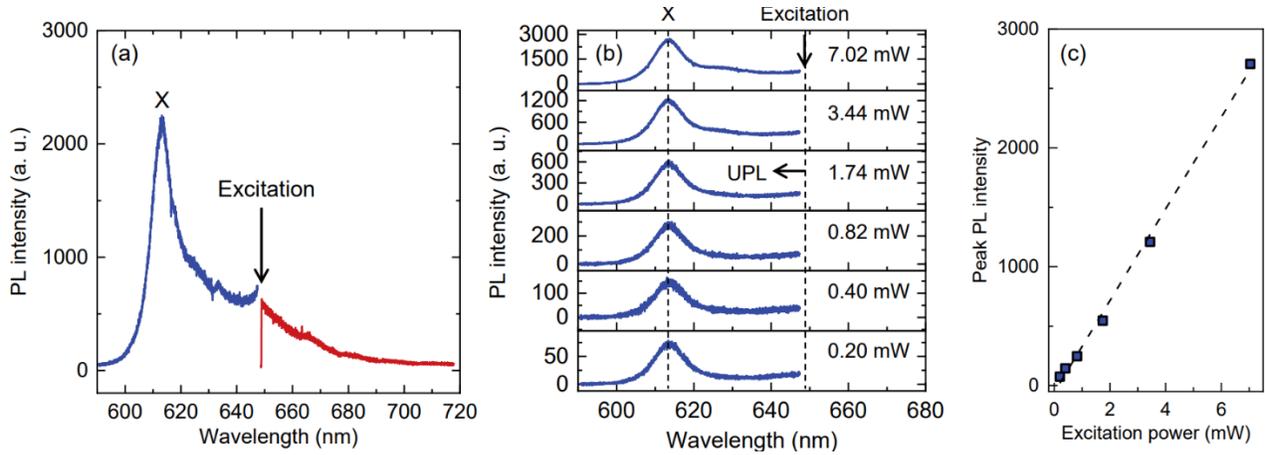

Figure S11. Up-conversion with continuous-wave (cw) laser excitation at 648 nm. (a) The up-converted and down-converted parts of the PL spectrum with an excitation power of 7.02 mW are shown by the blue and red lines, respectively. (b) The excitation-power-dependent up-conversion PL spectra. The corresponding excitation powers are listed for each spectrum. (c) The excitation-power-dependent peak PL intensity at X (613 nm). The black dashed line indicates the linear fit.

For up-conversion PL measurements, we did not aim at the exact center of the sample to observe X emission by D-band excitation. At the center of the NaBr-assisted WS$_2$ flake, the X-band emission was almost dissipated (as shown in Fig. 1f and Fig. S6). Therefore, we selected a region where both X and D emissions were observed, indicating a lower defect density compared to the center. As shown in Fig. S11a, the up-converted PL spectrum clearly shows a peak at 613 nm (2.0228 eV) corresponding to the X resonance, which is approximately 109 meV higher in photon energy compared to the excitation photon energy of 1.9136 eV (648 nm). Compared to 532 nm excitation, the peak wavelength of the X emission is blue-shifted by approximately 20 nm, which is attributed to a larger Stokes shift. Figure S11b shows excitation-power-dependent up-conversion PL spectra. There was no laser-power-dependent peak shift of the X emission band. The excitation-power-dependent intensity of the X emission was linear, as shown in Fig. S11c. In the pump–probe experiment, the X population due to D excitation was saturated (Fig. 6 in the main text) because the average photon flux of a cw laser ($\Phi_{cw} \approx 1.3 \times 13\ 10^{24}$ cm$^{-2}$ s$^{-1}$, considering a 648 nm wavelength, 7.02 mW excitation power, and 1.5 µm spot size) is about 1000 times lower than the peak photon flux of a pulse used in Fig. 6 in the main text ($\Phi_{pulse} \approx 1.7 \times 10^{27}$ cm$^{-2}$ s$^{-1}$, considering a 665 nm wavelength, 54.6 µJ/cm$^2$ fluence, and ~100 fs pulse width).

**Supplementary Figure S12**

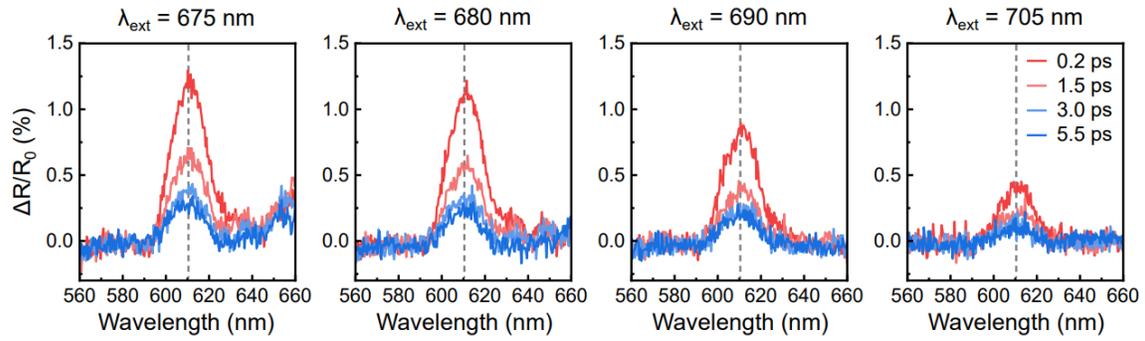

Figure S12. ΔR/R$_0$ spectra around the X band under D band-edge excitation, measured at different delays and excitation wavelengths as marked in the insets.

**Supplementary Figure S13**

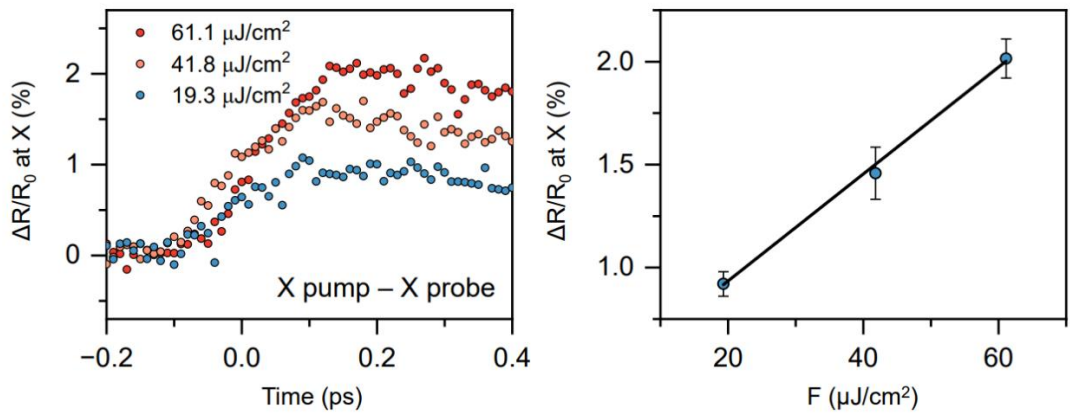

Figure S13. Pump fluence (F)-dependent ΔR/R$_0$ temporal evolution at the X-band-edge pump (595 nm) and X-band probe (612 nm), and their peak values as a function of F. The black curve indicates a linear fit.

**Supplementary Figure S14**

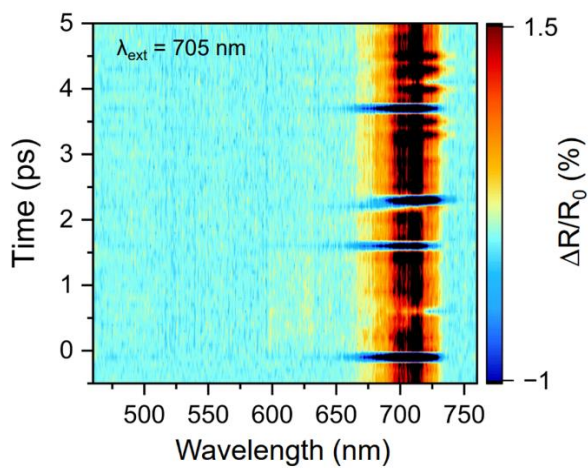

Figure S14. ΔR/R₀ map of sample #1 at a 705 nm pump wavelength.

**Supplementary Figure S15**

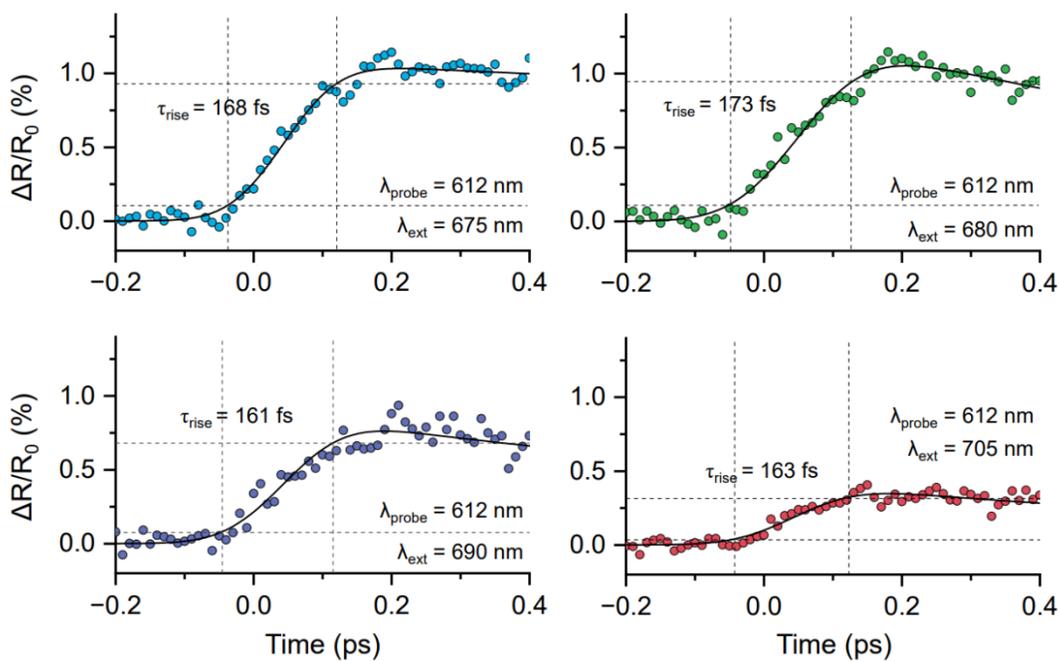

Figure S15. Pump-wavelength ($\lambda_{ext}$)-dependent temporal evolution of the $\Delta R/R_0$ signal at the X resonance (612 nm). The horizontal and vertical dashed lines indicate the 10% and 90% (of the peak value) positions used to estimate the apparent rise time.